\documentclass[aps,prd,preprint,draft,eqsecnum]{revtex4}
\usepackage{graphicx,epsf,amssymb}
\newcommand{\be}{\begin{equation}}
\newcommand{\ee}{\end{equation}}

\newcommand{\bea}{\begin{eqnarray}}
\newcommand{\eea}{\end{eqnarray}}
\newcommand{\Tr}{{\rm Tr}}
\newcommand{\nn}{\nonumber}

\newcommand{\dd}{\mbox{d}}
\newcommand{\gm}{\gamma}
\newcommand{\Gm}{\Gamma}
\newcommand{\lm}{\lambda}
\newcommand{\uz}{\underline{z}}
\newcommand{\bGm}{\bar{\Gamma}}
\newcommand{\sGm}{\Gamma^*}

\newif\ifdraft
\drafttrue
\newif\ifpreprint
\preprinttrue

\def\sect#1{section~{\ref{#1}}}
\def\app#1{appendix~{\ref{#1}}}
\def\App#1{Appendix~{\ref{#1}}}
\def\fig#1{fig.~{\ref{#1}}}

\def\Fig#1{Figure~{\ref{#1}}}

\def\tree{{\rm tree}}

\def\Tr{\, {\rm Tr}}
\def\SYM{MSYM}

\def\NeqFour{{N}=4}

\def\Nc{N_c}
\def\spa#1.#2{\left\langle#1\,#2\right\rangle}
\def\spb#1.#2{\left[#1\,#2\right]}
\def\sand#1.#2.#3{%
\left\langle\smash{#1}{\vphantom1}^{-}\right|{#2}%
\left|\smash{#3}{\vphantom1}^{-}\right\rangle}
\def\sandp#1.#2.#3{%
\left\langle\smash{#1}{\vphantom1}^{-}\right|{#2}%
\left|\smash{#3}{\vphantom1}^{+}\right\rangle}
\def\sandpp#1.#2.#3{%
\left\langle\smash{#1}{\vphantom1}^{+}\right|{#2}%
\left|\smash{#3}{\vphantom1}^{+}\right\rangle}
\def\sandpm#1.#2.#3{%
\left\langle\smash{#1}{\vphantom1}^{+}\right|{#2}%
\left|\smash{#3}{\vphantom1}^{-}\right\rangle}
\def\sandmp#1.#2.#3{%
\left\langle\smash{#1}{\vphantom1}^{-}\right|{#2}%
\left|\smash{#3}{\vphantom1}^{+}\right\rangle}
\def\sandmm#1.#2.#3{%
\left\langle\smash{#1}{\vphantom1}^{-}\right|{#2}%
\left|\smash{#3}{\vphantom1}^{-}\right\rangle}
\def\spab#1.#2.#3{\sandmm#1.#2.#3}
\def\spba#1.#2.#3{\sandpp#1.#2.#3}
\def\spaa#1.#2.#3.#4{\sandmp#1.{#2#3}.#4}
\def\spbb#1.#2.#3.#4{\sandpm#1.{#2#3}.#4}
\def\spash#1.#2{\spa{\smash{#1}}.{\smash{#2}}}

\newbox\charbox
\newbox\slabox
\def\s#1{{      
        \setbox\charbox=\hbox{$#1$}
        \setbox\slabox=\hbox{$/$}
        \dimen\charbox=\ht\slabox
        \advance\dimen\charbox by -\dp\slabox
        \advance\dimen\charbox by -\ht\charbox
        \advance\dimen\charbox by \dp\charbox
        \divide\dimen\charbox by 2
        \raise-\dimen\charbox\hbox to \wd\charbox{\hss/\hss}
        \llap{$#1$}
}}

\def\eqn#1{eq.~(\ref{#1})}
\def\Eqn#1{Equation~(\ref{#1})}
\def\eqns#1#2{eqs.~(\ref{#1}) and~(\ref{#2})}
\def\Eqns#1#2{Eqs.~(\ref{#1}) and~(\ref{#2})}
\def\qb{{\overline {\kern-0.7pt q\kern -0.7pt}}}

\def\e{\epsilon}
\def\ep{\epsilon}
\def\eps{\epsilon}

\def\tn#1#2{t^{[#1]}_{#2}}

\def\li#1{{\mathop{\rm Li}\nolimits}_#1}
\def\Li{\mathop{\rm Li}\nolimits}
\def\Split{\mathop{\rm Split}\nolimits}
\def\tree{{(0)}}
\def\Lloop{{(L)}}
\def\lloop{{(l)}}
\def\oneloop{{(1)}}
\def\twoloop{{(2)}}
\def\threeloop{{(3)}}

\def\cgh{\hat c_\Gamma}
\def\Ord{{\cal O}}
\def\mod{\mathop{\rm mod}\nolimits}

\def\H#1{H_{#1}}

\def\sandp#1.#2.#3{%
\left\langle\smash{#1}{\vphantom1}^{+}\right|{#2}%
\left|\smash{#3}{\vphantom1}^{+}\right\rangle}

\newbox\ourfigbox
\def\SizedFigureWithCaption#1#2#3{%
\setbox\ourfigbox
  \hbox{\hss\epsfxsize #1 \epsfbox{#2}\hss}
\hbox to \wd\ourfigbox{\vbox{\noindent\copy\ourfigbox\break
\vskip -6mm      \hbox to \wd\ourfigbox{\hss#3\hss}}}
}

\def\spa#1.#2{\left\langle#1\,#2\right\rangle}
\def\spb#1.#2{\left[#1\,#2\right]}
\def\lor#1.#2{\left(#1\,#2\right)}
\def\sand#1.#2.#3{%
\left\langle\smash{#1}{\vphantom1}^{-}\right|{#2}%
\left|\smash{#3}{\vphantom1}^{-}\right\rangle}
\def\sandpp#1.#2.#3{%
\left\langle\smash{#1}{\vphantom1}^{+}\right|{#2}%
\left|\smash{#3}{\vphantom1}^{+}\right\rangle}
\def\sandpm#1.#2.#3{%
\left\langle\smash{#1}{\vphantom1}^{+}\right|{#2}%
\left|\smash{#3}{\vphantom1}^{-}\right\rangle}
\def\sandmp#1.#2.#3{%
\left\langle\smash{#1}{\vphantom1}^{-}\right|{#2}%
\left|\smash{#3}{\vphantom1}^{+}\right\rangle}

\begin{document}
\hfuzz 10 pt


\ifpreprint
\noindent
 UCLA/05/TEP/14
\hfill SLAC--PUB--11210
\fi

\title{Iteration of Planar Amplitudes in\\
Maximally Supersymmetric Yang-Mills Theory\\
at Three Loops and Beyond}%

\author{Zvi Bern}
\affiliation{{} Department of Physics and Astronomy, UCLA\\
\hbox{Los Angeles, CA 90095--1547, USA}
}

\author{Lance J. Dixon}
\affiliation{ Stanford Linear Accelerator Center \\
              Stanford University\\
             Stanford, CA 94309, USA
}

\author{Vladimir A. Smirnov}
\affiliation{Nuclear Physics Institute of Moscow State University\\
\hbox{Moscow 119992, Russia}
}

\date{May, 2005}

\begin{abstract}
We compute the leading-color (planar) three-loop four-point amplitude
of $\NeqFour$ supersymmetric Yang-Mills theory in $4-2\e$ dimensions,
as a Laurent expansion about $\e=0$ including the finite terms.  The
amplitude was constructed previously via the unitarity method, in
terms of two Feynman loop integrals, one of which has been evaluated
already.  Here we use the Mellin-Barnes integration technique to
evaluate the Laurent expansion of the second integral.  Strikingly,
the amplitude is expressible, through the finite terms, in terms of
the corresponding one- and two-loop amplitudes, which provides strong
evidence for a previous conjecture that higher-loop planar $\NeqFour$
amplitudes have an iterative structure.  The infrared singularities of
the amplitude agree with the predictions of Sterman and Tejeda-Yeomans
based on resummation.  Based on the four-point result and the
exponentiation of infrared singularities, we give an exponentiated
ansatz for the maximally helicity-violating $n$-point amplitudes to
all loop orders.  The $1/\e^2$ pole in the four-point amplitude
determines the soft, or cusp, anomalous dimension at three loops in
$\NeqFour$ supersymmetric Yang-Mills theory.  The result confirms a
prediction by Kotikov, Lipatov, Onishchenko and Velizhanin, which
utilizes the leading-twist anomalous dimensions in QCD computed by
Moch, Vermaseren and Vogt.  Following similar logic, we are able to
predict a term in the three-loop quark and gluon form factors in QCD.
\end{abstract}

\pacs{11.15.Bt, 11.25.Db, 11.25.Tq, 11.55.Bq, 12.38.Bx \hspace{1cm}}

\maketitle



\renewcommand{\thefootnote}{\arabic{footnote}}
\setcounter{footnote}{0}

\section{Introduction}
\label{IntroSection}

Maximally supersymmetric $\NeqFour$ Yang-Mills theory (\SYM) in four
dimensions has a number of remarkable properties.  There are good
reasons to believe that, in the 't~Hooft (planar) limit of a large number
of colors $N_c$, higher-loop orders are
surprisingly simple~\cite{Iterate2}.  
In particular, the anti-de Sitter/conformal field theory (AdS/CFT) 
correspondence suggests a simplicity in the perturbative expansion of
planar \SYM\ as the number of loops increases~\cite{Iterate2}.
The Maldacena conjecture~\cite{Maldacena} states that the planar 
limit of \SYM\ at strong coupling is dual to
weakly-coupled gravity in five-dimensional anti-de Sitter space.
Based on this conjecture, one might expect observables in the
strongly-coupled limit of \SYM\ to have a relatively simple form, 
due to the interpretation in terms of weakly-coupled gravity.  
On the other hand, the strong-coupling limit of a typical observable
receives contributions from infinitely many terms in the perturbative 
expansion, as well as non-perturbative effects.   
How might the perturbative series be organized to produce a simple
strong-coupling result?  
Some quantities are protected by supersymmetry ---
non-renormalization theorems lead to zeroes in the perturbative series,
which certainly can bring about this simplicity~\cite{BPS,DhokerTasi}.  
It has been less clear how the perturbative series for unprotected 
quantities might have the required 
simplicity~\cite{OtherAnomalousDim,Nontrivial,DhokerTasi}.  

One suggestion, confirmed through two loops for dimensionally-regulated 
on-shell scattering amplitudes, is that an iterative 
structure exists~\cite{Iterate2}, which may eventually allow the 
perturbative series to be resummed into a simple result.
In particular, the planar four-point two-loop
amplitude of \SYM\ was shown to be expressible in
terms of the corresponding one-loop amplitude.  
Roughly speaking (see \eqn{TwoLoopOneLoop} for the precise formula),
the two-loop amplitude is given by the square of the one-loop
amplitude, plus a term proportional to the one-loop amplitude evaluated
in a slightly different dimension, plus a constant.
This result was found using the two-loop integrand~\cite{BRY,BDDPR} 
obtained via the unitarity method~\cite{NeqFourOneLoop,Fusing,
UnitarityMachinery,OneLoopReview, TwoLoopSplitting}, and the Laurent
expansion in $\e = (4-d)/2$ of the associated two-loop planar box
integral~\cite{SmirnovDoubleBox}.  

On-shell loop amplitudes in massless gauge theory have severe 
infrared (IR) singularities, arising from soft and collinear 
loop momenta.  Regulated dimensionally, the singularities produce 
poles in the limit $\e\to0$, beginning at $\Ord(\e^{-2L})$
for an $L$-loop amplitude. 
The two-loop iterative relation holds from $\Ord(\e^{-4})$
through $\Ord(\e^0)$, but it does not hold at $\Ord(\e^1)$.  
This observation is consistent with intuition that a simple structure
need only exist near four dimensions~\cite{Iterate2}, 
where \SYM\ is a conformal theory, and where it should be dual 
to a gravity theory in anti-de Sitter space. 

Splitting amplitudes are functions governing the behavior
of scattering amplitudes as two momenta become collinear.
The two-loop splitting amplitude in \SYM\ has an 
iterative structure very similar to that of the four-point 
amplitude~\cite{Iterate2,TwoLoopSplitting}.  
Based on this structure, an iterative ansatz for the planar $n$-point 
two-loop amplitudes can also be constructed.  The ansatz
is very likely to be true for the maximally helicity violating (MHV)
amplitudes (those with two negative helicities and the rest positive)
because it ensures that these amplitudes have the correct
factorization behavior in all channels.  (For non-MHV amplitudes one
would also need to ensure that the structure of the multi-particle
poles is correct.)  

Amplitudes for scattering of on-shell massless quanta have considerable
practical relevance, in the applications of perturbative QCD
to collider physics.  At the perturbative level, \SYM\ is a close cousin 
of QCD, although its amplitudes have a much simpler analytic structure,
allowing their computation typically to precede the QCD result.  In fact,
the surprisingly simple structure of \SYM\ loop amplitudes has been
unfolding for quite a while, beginning with the superstring-based 
evaluation of the one-loop four-point amplitude by Green, Schwarz 
and Brink~\cite{GSB}.  Compact results for the $n$-point MHV
amplitudes~\cite{NeqFourOneLoop}, and for all helicity configurations 
at six points~\cite{Fusing}, were among the early applications
of the unitarity method of Dunbar, Kosower, and two of the 
authors~\cite{NeqFourOneLoop,Fusing,UnitarityMachinery,OneLoopReview,
TwoLoopSplitting}. Because the unitarity method builds amplitudes at
any loop order from on-shell lower-loop amplitudes, any simplicity
uncovered at the tree and one-loop levels should induce a
corresponding additional simplicity at higher-loop orders.  Indeed,
the simplicity observed in the multi-loop four-point \SYM\ loop
integrands (prior to performing loop integrations) was found in this
way~\cite{BRY,BDDPR}.

Witten has proposed a duality between \SYM\
and twistor string theory~\cite{WittenTopologicalString},
generalizing Nair's earlier description~\cite{Nair} of 
MHV tree amplitudes.   
This proposal, and the investigations it has stimulated into the 
structure of tree~\cite{NewTree} and one-loop~\cite{NewOneloop,NMHV} 
gauge theory scattering amplitudes, 
provide additional strong support for the notion that 
amplitudes --- particularly \SYM\ amplitudes --- 
should be remarkably simple.

These results, particularly the two-loop iterative relation, 
lead to the natural conjecture that an iterative structure should 
continue to hold for higher-loop planar \SYM\ amplitudes~\cite{Iterate2}.  
The purpose of this paper is to verify the conjecture at the level 
of the three-loop four-point amplitude, and to flesh out more of the 
likely structure beyond three loops.

The planar three-loop four-point \SYM\ amplitude was found 
in ref.~\cite{BRY} via the unitarity method, and expressed in terms of 
just two independent loop integrals.
To check for an iterative relation, we must first compute the expansion of
these two integrals around $\e=0$, from the most singular terms, 
$\Ord(\e^{-6})$, through the finite terms, $\Ord(\e^0)$. 
Fortunately, there has been much progress in multi-loop
integration over the past few 
years~\cite{SmirnovDoubleBox,SmirnovVeretin,LoopIntegrationAdvance}.
One of the two integrals we need, a three-loop ladder integral, 
was computed through finite order recently by one of the 
authors~\cite{SmirnovTripleBox}, using a multiple Mellin-Barnes 
(MB) representation.  In this paper, we present the expansion of the 
single remaining integral --- and thus the expansion of the 
three-loop amplitude --- through the finite terms.

We wish to compare this expression with the expansions 
of products of one- and two-loop amplitudes. 
For this purpose, we must expand the one- and two-loop 
amplitudes to $\Ord(\e^4)$ and $\Ord(\e^2)$ respectively,
which is two higher orders in $\e$ than was necessary at two loops.  
All of the expansions are given in terms of harmonic 
polylogarithms~\cite{HPL,HPL2}. 
We use identities to reduce the harmonic polylogarithms to an
independent basis set.
Taking into account intricate cancellations between the
different amplitude terms, we find that the planar three-loop four-point 
amplitude does indeed have a simple iterative structure
(see \eqn{ThreeLoopFourPtIteration}).  

To guide us toward the correct iterative relation,
we employed properties of the three-loop amplitude's IR 
singularities~\cite{StermanTY}, which must be respected
by any such relation.
In general, the IR singularities of loop amplitudes in
gauge theory can be represented in terms of universal operators, 
acting on the same scattering amplitudes evaluated at lower loop
order, as was first discussed at one and two 
loops~\cite{IROneLoop,CataniIR}.
These operators are related to the soft (or cusp) anomalous 
dimension and other quantities entering the Sudakov form 
factor~\cite{Sudakov,MagneaSterman}, as was clarified 
recently~\cite{StermanTY}.
The latter quantities play an important role in the resummation 
and exponentiation of large logarithms near kinematic boundaries, 
such as the threshold ($x\to1$) logarithms in deep inelastic 
scattering or the Drell-Yan 
process~\cite{SoftGluonSummation,MagneaSterman,EynckLaenenMagnea}. 

In other words, the IR divergence structure of loop amplitudes are
{\it a priori} predictable, up to sets of numbers ({\it e.g.} 
soft anomalous dimensions) that must be obtained by specific 
computations.  
Our four-point computation simultaneously provides 
a verification of the three-loop IR divergence formula~\cite{StermanTY}, 
and a direct determination of two of the numbers entering it, 
for planar \SYM:  the three-loop coefficients of the soft anomalous 
dimension and of the ${\cal G}$-function for the Sudakov form 
factor~\cite{MagneaSterman,StermanTY}.

The three-loop four-point iterative relation, 
combined with information about
how IR singularities exponentiate~\cite{MagneaSterman},
and the factorization properties used at two 
loops~\cite{Iterate2}, leads us to an exponentiated ansatz 
for the planar $n$-point MHV amplitudes at $L$ loops.  
This ansatz naturally produces each loop amplitude 
as an iteration of lower-loop amplitudes,
up to a set of constants which are as yet undetermined beyond three loops.
(Two rational numbers at three loops are also undetermined.)
By taking collinear limits of the ansatz, we obtain, as a by-product, 
an iterative ansatz for the $L$-loop splitting amplitudes of \SYM.

We use the universal form of the divergences to define IR-subtracted
finite remainder amplitudes. (Similar subtractions are made in 
perturbative QCD when constructing finite cross sections for 
infrared-safe observables.)  For our exponentiated ansatz,
the finite remainder at $L$ loops is strikingly
simple:  it is a polynomial of degree $L$ in the one-loop finite remainder.
This result applies directly to the finite remainder of the three-loop
four-point amplitude, for which it follows from actual computation,
not an ansatz.

Infrared singularities provide a link between the scattering
amplitudes computed here and the anomalous dimensions of 
gauge-invariant composite operators in \SYM, studied in the context
of the AdS/CFT
correspondence~\cite{OtherAnomalousDim,DhokerTasi,Integrable,MoreIntegrable}.
Specifically, at three loops, the coefficient of the $1/\e^2$ 
IR singularity is controlled by the high-spin, or soft, limit 
of the leading-twist anomalous dimensions~\cite{StermanTY}.
Equivalently, it appears in the $x\to1$ limit of the kernels
for evolving parton distributions $f_i(x,Q^2)$ in the scale $Q^2$.
The $x\to1$ limit of the splitting kernels corresponds to multiple
soft gluon emission, and is related to the soft (or cusp) 
anomalous dimension associated with a Wilson line~\cite{KM}.
The three-loop soft anomalous dimension in QCD has been computed by 
Moch, Vermaseren and Vogt as part of the heroic computation 
of the full leading-twist anomalous dimensions~\cite{MVV}.  
(The terms proportional to $N_f$ were computed earlier~\cite{SoftNf}.)

The QCD result has been carried over to \SYM\ by Kotikov, Lipatov,
Onishchenko and Velizhanin (KLOV)~\cite{KLOV}, 
using an inspired observation that
the \SYM\ results may be obtained from the ``leading-transcendentality'' 
contributions of QCD.  For the soft anomalous dimensions, which
are polynomials in the Riemann $\zeta$ values, $\zeta_n \equiv \zeta(n)$,
the degree of transcendentality is tallied by assigning the degree
$n$ to each $\zeta_n$.  
The KLOV observation applies to the anomalous dimensions for any spin $j$;
a similar accounting of harmonic sums $S_{\vec{m}}(j)$ is used to assign
transcendentality in that case. 
Very interestingly, the three-loop \SYM\ anomalous dimensions of KLOV
were confirmed by Staudacher~\cite{Staudacher} through spin $j=8$,
building on earlier work of 
Beisert, Kristjansen and Staudacher~\cite{MoreIntegrable} at $j=4$,
by assuming integrability and using a Bethe ansatz.
Our determination of the three-loop soft anomalous dimension in \SYM\ 
now provides an independent confirmation of the KLOV result
in the limit $j\to\infty$.

The iterative structure of \SYM\ is presumably tied to the issue of
integrability of the theory~\cite{Integrable,MoreIntegrable}.  
There has also been an interesting hint of a similar structure 
developing in the correlation functions of gauge-invariant composite
operators in \SYM~\cite{Schubert}; but its precise structure, 
if it exists in this case, has not yet been clarified.

This paper is organized as follows.  In \sect{ReviewSection} we review
known results for planar loop amplitudes in \SYM, focusing on the
construction of the three-loop integrand for the four-point amplitude.
The methods used to evaluate the two three-loop integrals are
described in \sect{TripleBoxSection}.  In \sect{IterationSection} we
describe the iterative relation for the three-loop four-point
amplitude.  Then we present an exponentiated ansatz which extends the
relation to $n$-point MHV amplitudes at an arbitrary number of loops.
We discuss the consistency of this ansatz with exponentiation of
infrared singularities.  The consistency of our ansatz under 
factorization onto kinematic poles, particularly the collinear limits,
is discussed in \sect{CollBehaviorSection}.  
In \sect{AnomalousDimensionSection} we relate the 
anomalous dimensions and Sudakov coefficients appearing in 
the $L$-loop amplitudes to previous work in QCD and \SYM.  
Our conclusions are given in \sect{ConclusionSection}.  
\App{HarmonicPolyLogAppendix}
summarizes properties of harmonic polylogarithms, while
\app{IntegralsAppendix} contains the results for all loop integrals
encountered in our calculation of the amplitudes.

\section{General structure of \SYM\ loop amplitudes}
\label{ReviewSection}

It is convenient to first color decompose the
amplitudes~\cite{TreeReview,OneLoopReview} in order to separate the color
from the kinematics.  In this paper we will
discuss only the leading-color planar contributions.  These terms
have the same color decomposition as tree amplitudes, up to
overall factors of the number of colors, $N_c$.  The leading-$N_c$
contributions to the $L$-loop $SU(N_c)$ gauge-theory $n$-point
amplitudes may be written in the color-decomposed form as,
\begin{eqnarray}
{\cal A}_n^{(L)} & = & g^{n-2}
 \Biggl[ { 2 e^{- \e \gamma} g^2 N_c \over (4\pi)^{2-\e} } \Biggr]^{L}
 \sum_{\rho}
\Tr( T^{a_{\rho(1)}} 
   \ldots T^{a_{\rho(n)}} )
               A_n^{(L)}(\rho(1), \rho(2), \ldots, \rho(n))\,,
\label{LeadingColorDecomposition}
\end{eqnarray}
where $\gamma$ is Euler's constant, and
the sum runs over non-cyclic permutations of the external legs.
In this expression we have suppressed the (all-outgoing) momenta $k_i$ 
and helicities $\lambda_i$, leaving only the index $i$ as a label.  This
decomposition holds for all particles in the gauge super-multiplet
which are all in the adjoint representation.  The advantage of this
form is that the color-ordered partial amplitudes $A_n$ are 
independent of the color factors, 
cleanly separating color and kinematics.  We will not discuss
the subleading-color contributions here because there does not appear
to be a simple iterative structure present for them~\cite{Iterate2}.

In general, loop amplitudes in massless gauge theory, including \SYM,
contain IR singularities.  This implies that a textbook definition of
the $S$-matrix with fixed numbers of elementary particles does not
exist.  To define an $S$-matrix in massless gauge theory, dimensional
regularization --- which explicitly breaks the conformal invariance
--- is commonly used.  Once the universal IR singularities are
subtracted, the four-dimensional limit of the remaining terms in the
amplitudes may then be taken.  In QCD, after combining real emission
and virtual contributions, these finite remainders are the quantities
entering into the computation of infrared-safe physical
observables~\cite{LeeNauenberg}.  It is worth noting that the
finite remainders should also be related to perturbative scattering
matrix elements for appropriate coherent states (see {\it e.g.}
ref.~\cite{CEIR}).  The IR singularities for \SYM\ that we discuss in
this paper are closely connected to those of QCD and are, in fact, a
subset of the QCD divergences.  As is typical in perturbative QCD, 
the $S$-matrix under discussion here is not the one for the 
true asymptotic states of the four-dimensional theory, but for 
elementary partons.

The unitarity
method~\cite{NeqFourOneLoop,Fusing,UnitarityMachinery,OneLoopReview,
TwoLoopSplitting} provides an efficient means to obtain the integrands
needed for constructing loop amplitudes.  In this approach, the
integrands for loop amplitudes are obtained directly from on-shell
tree amplitudes without resorting to an off-shell formalism. A key
advantage is that the building blocks used to obtain the amplitudes
are gauge invariant and posses simple properties under extended
supersymmetry, unlike Feynman diagrams.  (Implicit in this approach is
the use of a supersymmetric regulator, such as the four-dimensional
helicity (FDH) scheme~\cite{FDH}, a variation on dimensional
reduction (DR)~\cite{Siegel}.) The unitarity method derives its efficiency
from the ability to use simplified forms of tree amplitudes to produce
simplified loop integrands.

The unitarity method expresses the amplitude in terms of a set of loop
integrals.  Experience shows that such integrals can be evaluated in
terms of generalized polylogarithms.  At one loop a complete basis of
dimensionally regularized integral functions is
known~\cite{OneLoopIntegralBasis,NeqFourOneLoop,Fusing}, in general,
reducing the integration problem to that of determining coefficients
of the basis integrals.  For four-point amplitudes only a single
scalar box integral appears.  At two and higher loops an analogous
basis of integral functions is not known, and the integrals must be
evaluated case by case.  The two-loop massless planar double-box
integral has, however, been evaluated in ref.~\cite{SmirnovDoubleBox}
and is given in terms of harmonic polylogarithms~\cite{HPL,HPL2} through
$\Ord(\e^2)$ in \eqn{TwoLoopBoxValue} of
the second appendix.  One of the integrals appearing in
the three-loop four-point amplitude has also been previously 
evaluated~\cite{SmirnovTripleBox}, and is given in \eqn{TripleBoxValueA}.

The one-loop four-point amplitude in \SYM\ was first calculated by
taking the low energy limit of a superstring~\cite{GSB}.  After
scaling out a factor of the tree amplitude via,
\begin{equation}
M_n^\Lloop(\e) = A_n^\Lloop/A_n^\tree \,,
\label{RescaledLoopAmplitude}
\end{equation}
the result for the one-loop four-point amplitude is rather simple,
\begin{equation}
M_4^{\oneloop}(\e) =
- {1 \over 2} \, s t \, I_4^\oneloop(s,t)\,.
\label{OneLoopAmplitude}
\end{equation}
Here $I_4^\oneloop$ is the one-loop scalar box
integral, multiplied by a convenient normalization factor,
and is defined in \eqn{OneLoopBoxDef} of
\app{OneLoopBoxIntegralSubsection}. This box integral is identical to
the one encountered in scalar $\phi^3$ theory.  Its explicit value in
terms of harmonic polylogarithms is given through $\Ord(\eps^4)$ in
\eqn{ExplicitOneloopBox}. We keep the higher-order terms in $\eps$
because they will contribute when we write the three-loop amplitude
in terms of the one- and two-loop amplitudes.  The factor of $1/2$ 
in~\eqn{OneLoopAmplitude} is due to our normalization convention
for $A_n^\Lloop$, exposed in \eqn{LeadingColorDecomposition} where
a compensatory ``2'' appears in the brackets.  This convention follows
the QCD literature on two-loop scattering amplitudes 
(see {\it e.g.}, ref.~\cite{GGGGTwoLoop}).

%
\begin{figure}[t]
\centerline{\epsfxsize 4. truein \epsfbox{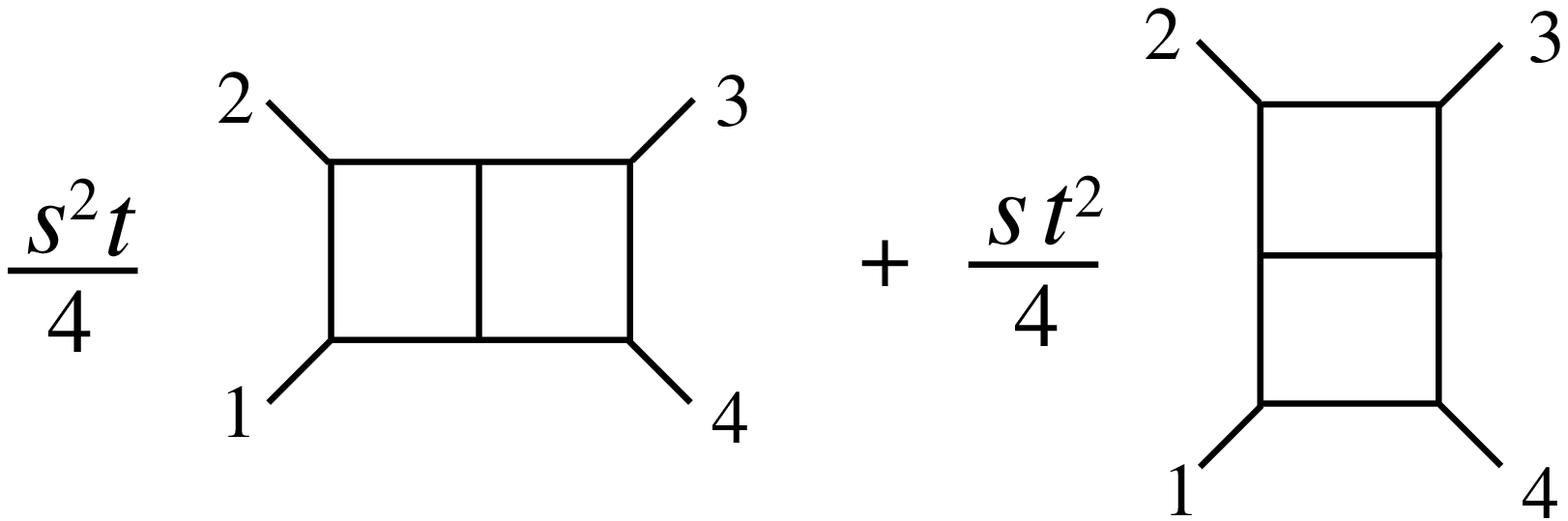}}
\caption{The result for the leading-color two-loop amplitude
in terms of scalar integral functions, given in~\eqn{TwoloopPlanarResult}.}
\label{TwoLoopResultFigure}
\end{figure}

The two-loop \SYM\ four-point amplitudes
were obtained in ref.~\cite{BRY} using the unitarity method,
with the result for the planar contribution,
\begin{equation}
M^\twoloop_4(\e) =
{1\over 4} s t \,
 \left( s \, I_4^\twoloop(s,t) + t \, I_4^\twoloop(t,s) \right) \,,
\label{TwoloopPlanarResult}
\end{equation}
which is schematically depicted in \fig{TwoLoopResultFigure}.  The
two-loop scalar integral $I_4^\twoloop$ is defined in
\eqn{TwoLoopBoxDef}.  The scalar double box integral
$I_4^\twoloop(s,t)$ was first evaluated through $\Ord(\e^0)$ in terms
of polylogarithms by one of the authors using multiple MB
representations~\cite{SmirnovDoubleBox}.  In \eqn{TwoLoopBoxValue}, we
give this integral through $\Ord(\e^2)$.  The higher-order terms in $\e$
are again needed because they will appear in our iterative
relation for the three-loop amplitude.
The result~(\ref{TwoloopPlanarResult}) has been confirmed using
the two-loop four-gluon QCD amplitude for helicities 
$({-}{-}{+}{+})$~\cite{GGGGTwoLoop},
which can be converted into the four-gluon amplitude in \SYM\ by 
adjusting the number and color of states circulating in
the loop~\cite{Iterate2}. 

%
\begin{figure}[t]
\centerline{\epsfxsize 4 truein \epsfbox{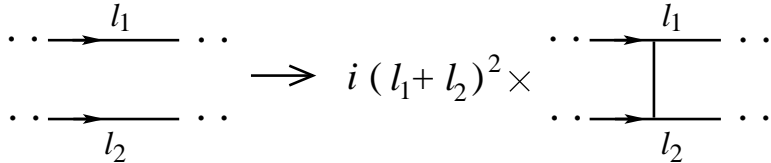}}
\caption{The rung insertion rule for generating higher-loop
integrands from lower loop ones, given in ref.~\cite{BRY}.}
\label{AddLineFigure}
\end{figure}

The original calculation~\cite{BRY} of the coefficients of the integrals in
\eqn{TwoloopPlanarResult} used iterated two-particle cuts, which are
known to be exact to all orders in $\e$ since they involve precisely
the same algebra used to obtain the one-loop
amplitude~(\ref{OneLoopAmplitude}).
Beyond two loops, an ansatz for the planar contributions to the
integrands was proposed in terms of a ``rung insertion
rule''~\cite{BRY,BDDPR}.  This ansatz was based on an analysis of two-
and three-particle cuts, as well as cuts with an arbitrary number of
intermediate states, but where the intermediate
helicities are restricted so that the amplitudes on either side of
the cut are MHV amplitudes.
At three loops, the planar integrals generated by 
the rung rule can be constructed using iterated two-particle cuts, 
so the ansatz is reasonably secure.
However, beyond three loops (and even at three loops for non-planar 
contributions) the rung rule generates diagram structures that cannot 
be obtained using iterated two-particle cuts.  
It is less certain that the rung rule gives the correct results 
for such contributions.  
There are also potential contributions coming from $(-2\e)$-dimensional 
parts of loop momenta, which have been dropped in the analysis of 
the three-particle and MHV cuts.
These contributions would need to be kept in order to prove rigorously
that the rung rule correctly gives all contributions. 

It is worth noting that while the integrand obtained from the 
rung insertion rule is only an ansatz, the results of this paper 
provide strong evidence that it is the complete answer,
at least for the planar contributions at three loops.  
As we shall discuss in \sect{IRConsistencySection},
the IR divergences of \eqn{ThreeLoopPlanarResult} are fully
consistent with the known form of the three-loop
IR divergences~\cite{StermanTY}.  
Moreover, the non-trivial cancellations
required by the iterative relations described in
\sect{IterationSection} imply that there are no missing pieces.

In any case, we use the rung rule as our starting point
for evaluating the planar three-loop \SYM\ amplitudes. According to
this rule one takes each diagram in the $L$-loop amplitude and
generates all the possible $(L+1)$-loop diagrams by inserting a new
leg between each possible pair of internal legs as shown in
\fig{AddLineFigure}.  From this set the diagrams which have triangle
or bubble subdiagrams are removed.  The new loop momentum is
integrated over, after including an additional factor of $i
(l_1+l_2)^2$ in the numerator, where $l_1$ and $l_2$ are the momenta
flowing through each of the legs to which the new line is joined, as
indicated in \fig{AddLineFigure}.  Each distinct $(L+1)$-loop
contribution should be counted once, even though it can be generated
in multiple ways.  (The contributions which correspond to identical
graphs but have different numerator factors should be counted as
distinct.)  The $(L+1)$-loop planar amplitude is then the sum of all
distinct $(L+1)$-loop diagrams.  The diagrams generated by the
iterated two-particle cuts have an amusing resemblance to Mondrian's
artwork; hence it is natural to call them ``Mondrian diagrams.''

%
\begin{figure}[t]
\centerline{\epsfxsize 6.0 truein \epsfbox{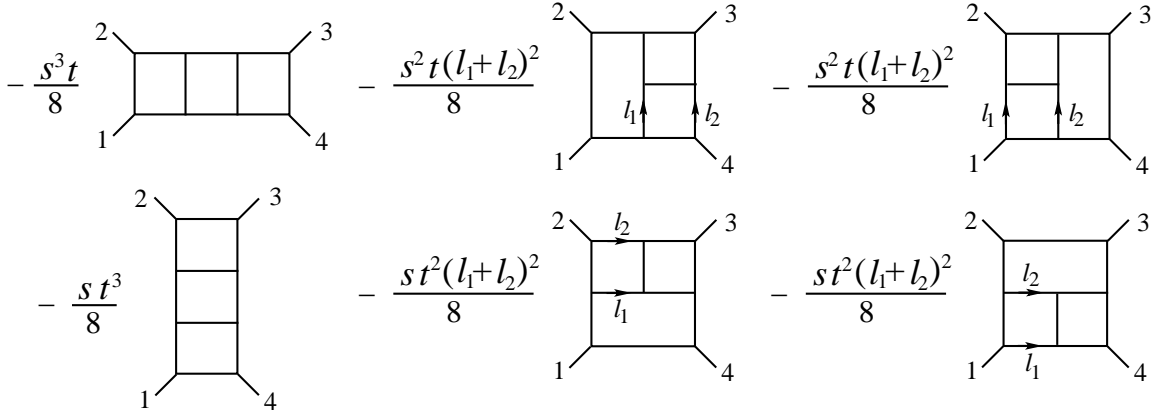}}
\caption{Mondrian diagrams for the three-loop four-point \SYM\ planar
amplitude given in \eqn{ThreeLoopPlanarResult}.  The second and third
diagrams have identical values, as do the fifth and sixth.  The factors
of $(l_1 + l_2)^2$ denote numerator factors appearing in the
integrals, where $l_1$ and $l_2$ are the momenta carried by the lines
marked by arrows.}
\label{ThreeLoopResultFigure}
\end{figure}

Applying this rule to the three-loop planar amplitude gives
the explicit form of the integrand~\cite{BRY},
\begin{equation}
M^\threeloop_4(\e) =
-{1\over8} s t \,
\Bigl(s^2 \, I_4^{\threeloop a}(s,t) +
      2 s \, I_4^{\threeloop b}(t,s) +
      t^2 \, I_4^{\threeloop a}(t,s) +
      2 t \, I_4^{\threeloop b}(s,t) \Bigr) \,.
\label{ThreeLoopPlanarResult}
\end{equation}
This integrand is depicted in
\fig{ThreeLoopResultFigure}~\footnote{The form in fig.~7
of ref.~\cite{BRY} is related to the one in \fig{ThreeLoopResultFigure}
by momentum conservation.}. The second and third integrals in the
figure are equal, as are the fifth and sixth, accounting for the
appearance of six diagrams in \fig{ThreeLoopResultFigure}, but only
four terms in \eqn{ThreeLoopPlanarResult}.  The integrals
$I_4^{\threeloop a}$ and $I_4^{\threeloop b}$ appearing in the
amplitude are defined in
\eqns{ThreeLoopIntegralA}{ThreeLoopIntegralB}.  The first of these
integrals has been evaluated in ref.~\cite{SmirnovTripleBox}.
The evaluation of the second integral is outlined in the next section.
The expansions of these integrals through
$\Ord(\e^0)$, in terms of harmonic polylogarithms, are presented in
\eqns{TripleBoxValueA}{TripleBoxValueB}.

\section{Evaluating triple boxes}
\label{TripleBoxSection}

%
\begin{figure}[t]
\centerline{\epsfxsize 4.5 truein \epsfbox{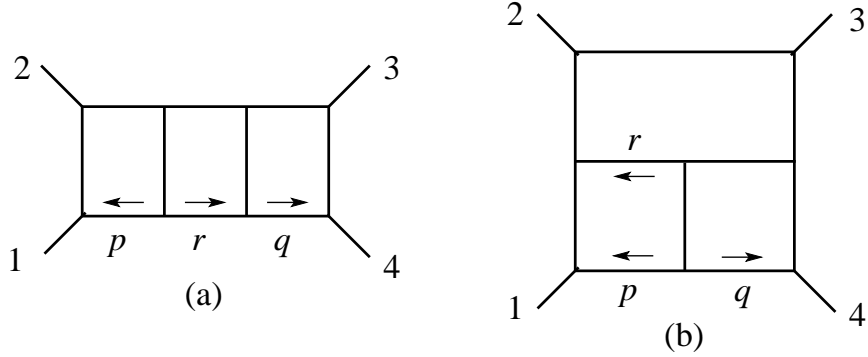}}
\caption{The two integrals appearing in the three-loop amplitude.
The ``ladder'' integral (a) has no factors in the numerator.
The ``tennis court'' integral (b) contains a factor of $(p+r)^2$ 
in the numerator.}
\label{TripleBoxFigure}
\end{figure}

The two three-loop integrals appearing in the four-point amplitude
(\ref{TwoloopPlanarResult}), and depicted in \fig{TripleBoxFigure} are
\begin{eqnarray}
I_4^{\threeloop a}(s,t) & = &
(-i e^{\e \gamma} \pi^{-d/2})^3
 \int
 {\dd^d p\,\dd^d r\,\dd^d q\, \over p^2 \, (p - k_1)^2 \,(p - k_1 - k_2)^2 }
\nonumber \\
&& \null \times
{1\over (p + r)^2 \,
    r^2 \, (q-r)^2 \, (r- k_3 - k_4)^2 \,
    q^2 \,  (q-k_4)^2 \, (q - k_3 - k_4)^2 }\,, \hskip 1 cm
\label{ThreeLoopIntegralA}
\end{eqnarray}
and
\begin{eqnarray}
I_4^{\threeloop b}(s,t) & = &
(-i e^{\e \gamma} \pi^{-d/2})^3
\int
{\dd^d p\,\dd^d r\,\dd^d q\,(p+r)^2 \over p^2 \, q^2 \, r^2 \,
(p - k_1)^2 \,(p +r - k_1)^2 }
\nonumber \\
&& \null \times
{1\over (p+r-k_1-k_2)^2 \, (p+r +k_4)^2 \, (q-k_4)^2
    (r+p+q)^2 \, (p+q)^2 } \,,
\label{ThreeLoopIntegralB}
\end{eqnarray}
where dimensional regularization with $d=4-2\ep$ is implied.

The ladder integral, $I_4^{\threeloop a}$, was evaluated 
in ref.~\cite{SmirnovTripleBox},
in a Laurent expansion in $\ep$ up to the finite part,
by means of the strategy based on the MB
representation which was suggested in
ref.~\cite{SmirnovDoubleBox} and applied for the evaluation of
the massless on-shell double boxes.
This strategy is presented in detail in Chapter~4 of
ref.~\cite{Buch}. Here its basic features are briefly
summarized.

The strategy starts with
the derivation of an appropriate multiple MB representation.
MB integrations are introduced in order to replace a sum of
terms raised to some power by their products raised to
certain powers, at the cost of having extra integrations:
\be
\frac{1}{(X+Y)^{\lm}} = \frac{1}{\Gm(\lm)}
\frac{1}{2\pi i}\int_{\beta-i \infty}^{\beta+i \infty} \dd z\,
\Gm(\lm+z) \Gm(-z) \frac{Y^z}{X^{\lm+z}}  \; ,
\label{MB}
\ee
where $-{\rm Re}\,\lambda < \beta < 0$.
The simplest possible way of introducing an MB integration
is to write down a massive propagator as a superposition of massless
ones. In complicated situations, one starts from Feynman or alpha
parameters and applies (\ref{MB}) to functions depending on
these parameters. Of course, it is natural to try to introduce
a minimal number of MB integrations. Anyway, after introducing
sufficiently many MB integrations, one can evaluate all
internal integrals over Feynman/alpha parameters in terms of gamma
functions and arrive at a multiple MB representation with an
integrand expressed in terms of gamma functions in the numerator
and denominator.

It turns out to be very convenient to derive a multiple MB representation
for loop-momentum integrals of a given class with general 
powers of the Feynman propagators.
Such a general derivation provides a lot of crucial checks and can then be
used for any integral of the given class.
Moreover, it provides unambiguous
prescriptions for choosing contours in MB integrals, where
the poles with $\Gm(\ldots-z)$ dependence are to the right of the
integration contour and the poles with $\Gm(\ldots+z)$ dependence
are to the left of it.

To evaluate a given Feynman integral represented in
terms of a multiple MB integral in an expansion in $\ep$ one needs
first to understand how poles in $\ep$ are generated.
A simple example is given by the product $\Gm(\ep+z)\Gm(-z)$ which
generates the singularity at $\ep \to 0$ because, in this
limit, there is no place for a contour to go between the first
left and right poles of these two gamma functions, at $z=-\ep$
and $z=0$, respectively.
To make the singular behavior in $\ep$ manifest one can
integrate instead over a new contour where the pole at 
$z=-\ep$ is to the right of the contour (for example, 
$\beta=-1$ in \eqn{MB}, where $\lambda=\ep$ is assumed to
have a small positive real part), plus a residue at this pole.
We refer to the integral over the new contour as ``changing the
nature'' of the first pole of $\Gm(\ep+z)$.
In complicated situations, singularities in $\ep$ are not visible
at once, after one of the MB integrations. To reveal them one uses
the general rule according to which the product
$\Gm(a+z)\Gm(b-z)$, with $a$ and $b$ depending on other
MB integration variables, generates, due to integration
over $z$, a singularity of the type $\Gm(a+b)$.

Thus, to reveal the singularities in $\ep$  one
analyzes various products of gamma functions in the numerator of a
given integrand, implying various orders of integration over
given MB variables. After such an analysis, one distinguishes
some key gamma functions which are responsible for the
generation of poles in $\ep$. Then one begins the procedure of
shifting contours and taking residues, starting from one of these
key gamma functions. After taking a residue, one arrives at
an integral with one integration less; one then performs an analysis of
the generation of singularities in $\ep$ in the same spirit as 
for the initial integral.  For the integral with the shifted contour,
one takes care of a second key gamma function in a similar way.
As a result of this procedure, one obtains a family of integrals
for which a Laurent expansion of the integrand is possible.
To evaluate these integrals expanded in $\ep$, up to some order,
one can use the second and the first Barnes lemma and their
corollaries.  A collection of relevant formulae are given in 
Appendix~D of ref.~\cite{Buch}.

The technique of multiple MB representation has turned out to 
be very successful, at least in the evaluation of four-point 
Feynman integrals with two or more loops and severe soft
and collinear singularities 
(see refs.~\cite{SmirnovDoubleBox,Tausk,TwoloopOffandMassive,SmirnovTripleBox}),
so that it is natural to apply it to the evaluation of 
the three-loop tennis-court integral~(\ref{ThreeLoopIntegralB}),
which is the only missing ingredient of our calculation.
Let us outline the main steps, following the strategy characterized above.

An appropriate MB representation can be derived straightforwardly,
in a way similar to the treatment of the ladder triple box 
integral~(\ref{ThreeLoopIntegralA}) in ref.~\cite{SmirnovTripleBox}.
Indeed, one can derive an auxiliary
MB representation for the double box with two legs off shell,
apply it to the double box subintegral in (\ref{ThreeLoopIntegralB}),
and then insert it into the well-known MB representation for the on-shell
box (see, {\it e.g.}, Chapter~4 of ref.~\cite{Buch}).
As a result, an eightfold MB representation can be
derived for the general diagram of \fig{TripleBoxFigure}b
with the
eleventh index corresponding to the numerator $[(p+r)^2]^{-a_{11}}$.
For our integral with the powers $a_1=\ldots=a_{10}=1$ and
$a_{11}=-1$, this gives
\bea
I_4^{\threeloop b} (s,t)= -\frac{ e^{3 \e \gamma} }{
\Gm(-2\ep)(-s)^{1+3\ep} t^2}&&
\nn \\ &&  \hspace*{-67mm}
\times
\frac{1}{(2\pi i)^8}
\int_{-i\infty}^{+i\infty}\ldots \int_{-i\infty}^{+i\infty}
\dd w\,\dd z_1 \left(\prod_{j=2}^7 \dd z_j \Gm(-z_j) \right)
\left(\frac{t}{s} \right)^{w}\Gamma(1 + 3 \ep + w)
\nn \\ &&  \hspace*{-67mm}\times
\frac{\Gamma(-3 \ep - w)\Gm(1 + z_1 + z_2 + z_3)
\Gm(-1 - \ep - z_1 - z_3)\Gm(1 + z_1 + z_4)}
{\Gm(1 - z_2) \Gm(1 - z_3)\Gm(1 - z_6)
\Gm(1 - 2 \ep + z_1 + z_2 + z_3)}
\nn \\ &&  \hspace*{-67mm}\times
\frac{\Gm(-1 - \ep -z_1 - z_2 - z_4)
 \Gm(2 + \ep + z_1 + z_2 + z_3 + z_4)}
{\Gm(-1 - 4 \ep - z_5)\Gm(1 - z_4 - z_7) \Gm(2 + 2 \ep + z_4 + z_5 + z_6 + z_7)}
\nn \\ &&  \hspace*{-67mm}\times
\Gm(-\ep + z_1 + z_3 - z_5)  \Gm(2 - w + z_5) \Gm(-1 + w - z_5 - z_6)
\nn \\ &&  \hspace*{-67mm}\times
\Gm(z_5 + z_7-z_1)
\Gm(1 + z_5 + z_6) \Gm(-1 + w - z_4 - z_5 - z_7)
\nn \\ &&  \hspace*{-67mm}\times
\Gm(-\ep + z_1 + z_2 - z_5 -z_6 - z_7)
\Gm(1 - \ep - w + z_4 + z_5 + z_6 + z_7)
\nn \\ &&  \hspace*{-67mm}\times
\Gm(1 + \ep - z_1 - z_2 - z_3 + z_5 + z_6 + z_7)
\; .
\label{3box2-8MB}
\eea
There is a factor of $\Gm(-2\ep)$
in the denominator, so that the integral is effectively sevenfold.

A preliminary analysis shows that the following two gamma
functions are crucial for the generation of poles in $\ep$:
\be
\Gm(-1 + w - z_5 - z_6) \Gm(-1 + w - z_4 - z_5 - z_7)\,.
\ee
The first decomposition of (\ref{3box2-8MB}) reduces to taking
residues and shifting contours with respect to the first
poles of these two functions. We obtain
\be
T \equiv I_4^{\threeloop b}=T_{00}+T_{01}+T_{10}+T_{11}\,.
\ee
The term $T_{01}$ denotes minus the residue at $z_7= -1 + w - z_4 - z_5$
and changing the nature of the first pole of
$\Gm(-1 + w - z_5 - z_6)$;
the term $T_{10}$ denotes minus the residue at $z_6= -1 + w - z_5$
and changing the nature of the first pole
of $\Gm(-1 + w - z_4 - z_5 - z_7)$;
the term $T_{11}$ corresponds to taking both residues; 
and $T_{00}$ refers to changing the nature of both poles 
under consideration.

For each of these four terms,
one proceeds further using the strategy of shifting contours and
taking residues.
One can arrive at contributions
which are labelled by sequences of gamma functions.
Let us denote by $\bGm(\ldots \pm\uz_i)$ taking the 
residue at the first pole
of this gamma function with respect to the variable $z_i$,
and by
$\sGm(\ldots \pm\uz_i)$ changing the nature of this
pole. If $\Gm(\ldots \pm\uz_i)$ participates then both variants are
implied.  If there is only one $z$-variable in an argument of a
gamma function then it is not underlined.
The contributions that start from order $\ep^1$ in the
Laurent expansion are not listed. So, for $T_{00}$, one can arrive
at the following eleven contributions:

\noindent $\{\bGm(-1-\ep-\uz_1-z_3),\bGm(-\ep+\uz_2),\bGm(-1-2\ep-z_5),\bGm(-2\ep+z_6),
\bGm(-\ep+z_3+\uz_7),$

\noindent
$\bGm(-\ep+z_3),\Gm(-2\ep+z_4)\}$,

\noindent $\{\bGm(-1-\ep-\uz_1-z_3),\bGm(-\ep+\uz_2),\bGm(-1-2\ep-z_5),\bGm(-2\ep+z_6),
\bGm(-\ep+z_3+\uz_7)$,

\noindent
$\sGm(-\ep+z_3),\Gm(-\ep-z_3+\uz_4)\}$,

\noindent
$\{\bGm(-1-\ep-\uz_1-z_3),\bGm(-\ep+\uz_2),\bGm(-1-2\ep-z_5),\bGm(-2\ep+z_6),$

\noindent
$\sGm(-\ep+z_3+\uz_7),\bGm(-\ep-z_3+\uz_4)\}$,

\noindent $\{\bGm(-1-\ep-\uz_1-z_3),\bGm(-\ep+\uz_2),\bGm(-1-2\ep-z_5),\sGm(-2\ep+z_6),
\bGm(-\ep+z_3+\uz_7),$

\noindent
$\bGm(-\ep+z_3),\bGm(-2\ep+z_4)\}$,

\noindent $\{\bGm(-1-\ep-\uz_1-z_3),\bGm(-\ep+\uz_2),\sGm(-1-2\ep-z_5),
\bGm(-\ep-z_3+\uz_4)\}$,

\noindent $\{\bGm(-1-\ep-\uz_1-z_3),\sGm(-\ep+\uz_2),\bGm(-1-2\ep-z_5),\bGm(-2\ep+z_6),
\bGm(-\ep+z_3+\uz_7),$

\noindent
$\bGm(-\ep+z_3),\Gm(-2\ep+z_4)\}$,

\noindent
$\{\bGm(-1-\ep-\uz_1-z_3),\sGm(-\ep+\uz_2),\bGm(-1-2\ep-z_5),\sGm(-2\ep+z_6),
\bGm(-\ep+z_3+\uz_7),$

\noindent
$\bGm(-\ep+z_3),\bGm(-2\ep+z_4)\}$,

\noindent $\{\sGm(-1-\ep-\uz_1-z_3),\bGm(-\ep+z_1+z_3-\uz_5),
\bGm(-\ep+z_3+\uz_7),\bGm(-\ep+z_3)\}$. \\
The rest of the 203 contributions present in $T_{01}+T_{10}+T_{11}$
can be described in a similar way.

The final result for (\ref{ThreeLoopIntegralB}) is presented in
\eqn{TripleBoxValueB} of
\app{ThreeLoopBoxIntegralSubsection}.
The evaluation of this integral has turned out to be rather
intricate. The level of complexity is roughly five times
the corresponding complexity of the ladder triple box.
Therefore, systematic checks are quite desirable.
A powerful independent check can be provided by evaluating 
the leading orders of the asymptotic behavior in some limit.
Indeed, such checks were essential in previous calculations
--- see refs.~\cite{SmirnovDoubleBox,TwoloopOffandMassive,SmirnovTripleBox}.
Here we shall outline an independent evaluation of the dominant 
terms in \eqn{TripleBoxValueB} in the limit $s/t \to 0$.     

The limit $s/t \to 0$ is of the Regge type which is
typical of Minkowski space. Hence the well-known prescriptions
for limits typical of Euclidean space,
written in terms of a sum over subgraphs of a certain class
(see refs.~\cite{aeEucl,Book}), are not applicable here.
However, one can use more general prescriptions formulated
in terms of the so-called strategy of expansion
by regions \cite{BS,SR,Book}.  This approach is universal and
applicable for expanding any given Feynman integral in any asymptotic regime.

An essential point of this strategy is to reveal regions
in the space of the loop momenta which generate non-zero
contributions. A given region is characterized by some
relations between components of the loop momenta.
In particular, in the case of our limit $s/t \to 0$,
in the region where all the loop momenta are hard,
all the components of the loop momenta are of order $\sqrt{t}$.
It turns out that the most typical regions relevant to the
Regge and Sudakov limits are 1-collinear (1c) and 2-collinear (2c)
regions.  (Here ``$a$-collinear'' means that an appropriate loop 
momentum is collinear with external leg $a$.)
The crucial part of the strategy of expansion by regions
\cite{BS} is to expand the integrand in a Taylor series
in parameters which are small in a given region and then
{\em extend} the integration to the whole space of the loop
momenta, {\it i.e.,} {\em forget} about the initial region.
Another prescription of this strategy is to put to zero
any integral without scale (even if it is not regularized by
dimensional regularization).

In the case of the ladder triple box (\ref{ThreeLoopIntegralA}), 
in the Regge limit $t/s \to 0$,
only the (1c-1c-1c) and (2c-2c-2c) regions participate in the 
leading power-law behavior~\cite{ThreeboxRegge}.

For the tennis-court integral~(\ref{ThreeLoopIntegralB}),
the evaluation procedure outlined above is formulated in such a way 
that the leading terms of the expansion at $s/t \to 0$ 
can be clearly distinguished.
All these terms arise after taking residues with respect to the
variable $w$ at $w=0$ or $w=\ep$.
It turns out that only one contribution to the result
(\ref{TripleBoxValueB}) arises after taking a residue at
$w=\ep$. It involves no integration ({\it i.e.}, it is obtained from
(\ref{3box2-8MB}) by taking consecutively eight residues), so that
it can be expressed in terms of gamma functions for general 
values of $\ep$:
\bea
 I_4^{\threeloop b,c-c-us} (s,t)=
\frac{ e^{3 \e \gamma} }
{(-s)^{1+4\ep}\, (-t)^{2-\ep}\,\ep}
 \Gm(-\ep)^3 \Gm(\ep)^2 \Gm(1 + 2\ep)^2\,.
\label{1uc-2c-2c}
\eea

It turns out that this term is nothing but the
(1c-4c-us) contribution within the expansion by regions.
It is generated by the region where the momentum
of the line between the external vertices with momenta
$k_2$ and $k_3$ is considered ultrasoft (us), the loop momentum
of the left box subgraph is considered 1-collinear and
the loop momentum of the right box subgraph is considered
4-collinear --- see \fig{TripleBoxFigure}(b).
(Details of the expansion in the Sudakov and
Regge limits within the expansion by regions can be found
in ref.~\cite{SR} and Chapter~8 of ref.~\cite{Book}.)

The rest of the contributions to the leading power-law behavior
in the limit $s/t \to 0$ correspond to taking residues at $w=0$.
They can be identified as the sum of the (1c-1c-1c) and (4c-4c-4c)
contributions.
The (4c-4c-4c) contribution can be represented by the following 
fivefold MB integral:
\bea
I_4^{\threeloop b,4c-4c-4c} (s,t)=
-\frac{ e^{3 \e \gamma} }
{(-s)^{1+3\ep+x_2}\, (-t)^{2+x_1}}
\frac{\Gm(1 + 3 \ep +x_2) \Gm(-3 \ep -x_2)}{\Gm(-2 \ep - x_1)
\Gm(1 + x_1) \Gm(1 + x_2)}
&& \nn \\ &&  \hspace*{-130mm}\times
\frac{1}{(2\pi i)^5}
\int_{-i\infty}^{+i\infty}\ldots \int_{-i\infty}^{+i\infty}
\prod_{j=1}^5 \dd z_j
\frac{\Gm(1 + z_1)\Gm(1 + \ep + z_2 + z_3) \Gm(-z_2) \Gm(-\ep - z_3) \Gm(-z_3)}
{\Gm(1 - z_2) \Gm(1 - z_3)}
\nn \\ &&  \hspace*{-130mm}\times
\frac{ \Gm(-1 - \ep - z_1 - z_2)
\Gm(1 - x_1 + z_1 + z_2 + z_3) \Gm(-1 + x_1 - z_1 - z_4)}
{\Gm(1 - 2 \ep - x_1 - x_2 + z_1 + z_2 + z_3) \Gm(-2 - 4 \ep - x_2 - z_1 -z_4)}
\nn \\ &&  \hspace*{-130mm}\times
\frac{\Gm(1 + z_4)
\Gm(\ep + x_2 - z_2 - z_3 - z_5)
\Gm(-2 + \ep + x_2 - z_1 - z_2 - z_3 - z_4 -z_5)}
{\Gm(1 + \ep + x_2 - z_2 - z_3 - z_5) \Gm(1 + \ep + z_2 + z_3 + z_5)}
\nn \\ &&  \hspace*{-130mm}\times
\Gm(1 + z_5)\Gm(-1 - \ep - x_2 + z_2 - z_4)
\Gm(-2 \ep - x_2 + z_2 + z_3 + z_5)
\nn \\ &&  \hspace*{-130mm}\times
\Gm(2 - \ep - x_1 + z_1 + z_2 + z_3 + z_4 + z_5) \Gm(-z_2 - z_5)
\,.
\label{2c2c2c}
\eea
An auxiliary analytic regularization, by means of $x_1$ and $x_2$, is
introduced into the powers of the propagators with the momenta
$p-k_1$ and $q-k_4$.
The (1c-1c-1c) contribution can be obtained from (\ref{2c2c2c}) by the
permutation $x_1 \leftrightarrow x_2$.
Each of the two (c-c-c) contributions is singular at $x_1,x_2
\to 0$. The singularities are however cancelled in the sum.
It is reasonable to start by revealing  this singularity.
One can observe that it appears due to the product
\be
\Gm(2 - \ep - x_1 + z_1 + z_2 + z_3 + z_4 + z_5)
\Gm(-2 + \ep + x_2 - z_1 - z_2 - z_3 - z_4 - z_5)\,,
\ee
where the sum of the arguments of these gamma functions is
$x_2-x_1$.

So, the starting point is to take minus the residue at
$z_5= -2 + \ep + x_2 - z_1 - z_2 - z_3 - z_4$ and shift the
integration contour correspondingly.
The value of the residue is then symmetrized by $x_1 \leftrightarrow
x_2$. This sum leads, in the limit $x_1,x_2 \to 0$, to the
following fourfold MB integral:
\bea
I_4^{\threeloop b,c-c-c,res} (s,t)=
-\frac{ e^{3 \e \gamma} }
{(-s)^{1+3\ep}\, t^2}
\frac{\Gm(-3 \ep) \Gm(1 + 3 \ep) }{\Gm(-2 \ep)}
&& \nn \\ &&  \hspace*{-90mm}\times
\frac{1}{(2\pi i)^4}
\int_{-i\infty}^{+i\infty}\ldots \int_{-i\infty}^{+i\infty}
\prod_{j=1}^4 \dd z_j
\frac{\Gm(1 + z_1) \Gm(-1 - \ep - z_1 - z_2)
\Gm(-z_2) \Gm(-\ep -z_3) \Gm(-z_3) }
{ \Gm(1 - z_2) \Gm(1 - z_3)
\Gm(1 - 2 \ep + z_1 + z_2 + z_3)}
\nn \\ &&  \hspace*{-90mm}\times
\frac{\Gm(1 + z_1 + z_2 + z_3)\Gm(1 + \ep + z_2 + z_3)
\Gm(-1 - \ep + z_2 - z_4) \Gm(-1 - z_1 - z_4)\Gm(1 + z_4)}
{ \Gm(-2 - 4 \ep - z_1 - z_4)
\Gm(-1 + 2 \ep - z_1 - z_4) \Gm(3 + z_1 + z_4)}
\nn \\ &&  \hspace*{-90mm}\times
\Gm(-1 + \ep - z_1 - z_2 - z_3 - z_4)
\Gm(2 + z_1 + z_4)\Gm(-2 - \ep - z_1 -z_4)
\Gm(2 - \ep + z_1 + z_3 + z_4)
\nn \\ &&  \hspace*{-90mm}\times
\left[2 \gm + L + \psi(-3 \ep) +
\psi(-2 \ep) - \psi(1 + 3 \ep) -
\psi(1 + z_1 + z_2 + z_3) + \psi(-1 - z_1 - z_4)
\right.
\nn \\ &&  \hspace*{-90mm}
-\psi(-2 - 4 \ep - z_1 - z_4) +
\psi(-1 + 2 \ep - z_1 - z_4) +
\psi(-1 - \ep + z_2 - z_4)
\nn \\ &&  \hspace*{-90mm}
\left.
- \psi(-1 + \ep - z_1 - z_2 - z_3 - z_4) +
\psi(2 - \ep + z_1 + z_3 + z_4)\right]
\,,
\label{2c2c2cRes}
\eea
where $L=\ln(s/t)$.

In the integral over the shifted contour in $z_5$, one can set
$x_1=x_2=0$ to obtain the following fivefold integral:
\bea
I_4^{\threeloop b,c-c-c,int} (s,t)=
-\frac{2 \, e^{3 \e \gamma} }
{(-s)^{1+3\ep}\, (-t)^{2}}
\frac{\Gm(1 + 3 \ep) \Gm(-3 \ep)}{\Gm(-2 \ep)}
&& \nn \\ &&  \hspace*{-100mm}\times
\frac{1}{(2\pi i)^5}
\int_{-i\infty}^{+i\infty}\ldots \int_{-i\infty}^{+i\infty}
\prod_{j=1}^5 \dd z_j
\frac{\Gm(1 + z_1)\Gm(1 + \ep + z_2 + z_3) \Gm(-z_2) \Gm(-\ep - z_3) \Gm(-z_3)}
{\Gm(1 - z_2) \Gm(1 - z_3)}
\nn \\ &&  \hspace*{-100mm}\times
\frac{ \Gm(-1 - \ep - z_1 - z_2)
\Gm(1 + z_1 + z_2 + z_3) \Gm(-1 - z_1 - z_4)}
{\Gm(1 - 2 \ep + z_1 + z_2 + z_3) \Gm(-2 - 4 \ep - z_1 -z_4)}
\nn \\ &&  \hspace*{-100mm}\times
\frac{\Gm(1 + z_4)
\Gm(\ep - z_2 - z_3 - z_5)
\Gm^*(-2 + \ep  - z_1 - z_2 - z_3 - z_4 -z_5)}
{\Gm(1 + \ep - z_2 - z_3 - z_5) \Gm(1 + \ep + z_2 + z_3 + z_5)}
\nn \\ &&  \hspace*{-100mm}\times
\Gm(1 + z_5)\Gm(-1 - \ep  + z_2 - z_4)
\Gm(-2 \ep  + z_2 + z_3 + z_5)
\nn \\ &&  \hspace*{-100mm}\times
\Gm(2 - \ep + z_1 + z_2 + z_3 + z_4 + z_5) \Gm(-z_2 - z_5)
\,,
\label{ccc-shift}
\eea
where the asterisk on one of the gamma functions implies that
the first pole is considered to be of the opposite nature.

The evaluation of (\ref{2c2c2cRes}) and (\ref{ccc-shift}), in
an expansion in $\ep$,
is then performed according to the strategy characterized above.
After the resolution
of the singularities in $\ep$ one obtains 60 contributions where
an expansion of the integrand in $\ep$ becomes possible.
Eventually, one reproduces the following leading asymptotic behavior:
\def\hs{\hspace*{5mm}}
\bea
I_4^{\threeloop b} (s,t) &=&
-\frac{ 1 }
{(-s)^{1+3\ep}\, t^2}
 \nn \\ && \times
\left\{
{16\over 9}{1\over \ep^6} + {13\over 6} L{}{1\over \ep^5}
+ \left[{1\over 2}L^2 - {19\over 12} \pi^2\right]{1\over \ep^4}
+ \left[-{1\over 6}L^3 - {67\over 72}\pi^2 L
- {241\over 18} \zeta_3\right]{1\over \ep^3}
\right.
\nn \\ &&  \hs
+ \left[{1\over 24}L^4 + {13\over 24} \pi^2 L^2
- {67\over 6} \zeta_3 L - {19\over 6480} \pi^4\right]{1\over \ep^2}
\nn \\ &&  \hs
+ \left[- {1\over 120} L^5 - {13\over 72} \pi^2 L^3 
- {5\over 2} \zeta_3 L^2 
- {6523\over 8640} \pi^4 L
+ {1385\over 216} \pi^2 \zeta_3
- {1129\over 10} \zeta_5\right]{1\over \ep}
\nn \\ &&  \hs
+ {1\over 720} L^6 + {13\over 288} \pi^2 L^4 
+ {5\over 6}  \zeta_3 L^3 
+ {331\over 960} \pi^4 L^2
+ \biggl( {317\over 72} \pi^2 \zeta_3
- {1203\over 10} \zeta_5 \biggr) L 
\nn \\ &&  \hs
\left.
- {180631\over 3265920} \pi^6
- {163\over 6} \zeta_3^2 + \Ord\left(\frac{s}{t} \right)
\right\}
\,.
\label{3box2LO}
\eea
To compare \eqn{3box2LO}
with the complete result~(\ref{TripleBoxValueB}),
we use transformation formulae such as \eqn{invertx}
to invert the arguments of the harmonic polylogarithms.
The resulting quantities $H_{a_1 a_2 \ldots a_n}(1/x)$ with 
$a_n=1$ vanish as $x\to\infty$.  Logarithms are generated
by the transformation; these logarithms combine with the ones
already manifest in \eqn{TripleBoxValueB}, yielding an expression
in complete agreement with \eqn{3box2LO}.

\section{Iterative structure of amplitudes}
\label{IterationSection}

The iterative structure of the four-point \SYM\ amplitude
found at two loops is~\cite{Iterate2,TwoLoopSplitting},
\begin{equation}
M_4^{\twoloop}(\e) 
= {1 \over 2} \Bigl(M_4^{\oneloop}(\e) \Bigr)^2
 + f^\twoloop(\e) \, M_4^{\oneloop}(2\e) + C^{(2)}
 + \Ord(\e)\,,
\label{TwoLoopOneLoop}
\end{equation}
where
\begin{equation}
f^\twoloop(\e) = - (\zeta_2 + \zeta_3 \e + \zeta_4 \e^2 + \cdots) \,,
\label{f2def}
\end{equation}
and the constant $C^{(2)}$ is given by
\begin{equation}
 C^{(2)} =  - {1\over 2} \zeta_2^2 \,.
\label{C2def}
\end{equation}
This relation can be verified by inserting the  
expansion~(\ref{TwoLoopPDBResult}) for the planar double-box 
integral $I_4^\twoloop$
into \eqn{TwoloopPlanarResult} for $M^\twoloop_4(\e)$,
and the expansion~(\ref{OneLoopBoxResult}) for the one-loop box 
integral $I_4^\oneloop$ into \eqn{OneLoopAmplitude} for $M^\oneloop_4(\e)$.  
Up through the finite terms in $\e$, only harmonic polylogarithms 
with weights up to four are encountered 
(see~\app{HarmonicPolyLogAppendix}).
These functions can all be written in terms of ordinary 
polylogarithms if desired.
Non-trivial cancellations between weight-four polylogarithms
are needed to obtain \eqn{TwoLoopOneLoop}, 
strongly suggesting that the relation is not accidental and leading
to the conjecture that an iterative structure exists in the amplitudes
to all loop orders.

As mentioned in the introduction, the 
relationship~(\ref{TwoLoopOneLoop}) is valid only through $\Ord(\e^0)$, 
{\it i.e.} near four dimensions, where \SYM\ is conformal 
and the AdS/CFT correspondence should be applicable.
At $\Ord(\e^1)$, the difference between the left- and right-hand sides
is an unenlightening combination of weight-five harmonic polylogarithms,
not a simple constant.

In order to search for a relation similar to \eqn{TwoLoopOneLoop}
at three loops, we have substituted the values of the integrals
$I_4^{\threeloop a}$ and $I_4^{\threeloop b}$,
given in \eqns{ThreeLoopaResult}{ThreeLoopbResult} respectively,
into \eqn{ThreeLoopPlanarResult} for $M^\threeloop_4(\e)$.
We have also used the $\e$-expansions of the 
one- and two-loop amplitudes through $\Ord(\e^4)$ and $\Ord(\e^2)$ 
respectively, two further orders than required for the 
two-loop relation~(\ref{TwoLoopOneLoop}).
(We {\it cannot} use \eqn{TwoLoopOneLoop} to replace $M_4^\twoloop$ 
with $M_4^\oneloop$, because that equation is valid only through $\Ord(\e^0)$.)
Thus we have obtained a representation
of the amplitudes in terms of harmonic polylogarithms~\cite{HPL,HPL2}
with weights up to six.  Because harmonic polylogarithms
with arguments equal to $-t/s$ and $-s/t$ both appear,
we need to employ identities which invert the argument, of the 
type outlined in \app{HarmonicPolyLogAppendix}.

Motivated also by the structure of the three-loop
IR divergences described in ref.~\cite{StermanTY},
we have found the following iterative relation for the three-loop
four-point amplitude,
\begin{eqnarray}
M_4^\threeloop(\e) &=& - {1\over 3} \Bigl[M_4^\oneloop(\e)\Bigr]^3
            + M_4^\oneloop(\e)\, M_4^\twoloop(\e)
            + f^\threeloop(\e) \, M_4^\oneloop (3\,\e) + C^{(3)}
            + \Ord(\e) \,,
\label{ThreeLoopFourPtIteration}
\end{eqnarray}
where
\begin{equation}
f^\threeloop(\e) = {11\over 2} \, \zeta_4
                 + \e (6 \zeta_5 + 5 \zeta_2 \zeta_3 ) +
   \e^2 (c_1 \zeta_6 +  c_2\zeta_3^2) \,,
\label{f3def}
\end{equation}
and the constant $C^{(3)}$ is given by
\begin{equation}
 C^{(3)} =  \biggl( {341\over 216} \, + {2\over 9} c_1 \biggl) \zeta_6
            + \biggl( - {17\over 9} + {2\over 9} c_2 \biggr)\zeta_3^2\,.
\label{C3def}
\end{equation}
The constants $c_1$ and $c_2$ are expected to be rational numbers.
They do not contribute to the right-hand side of 
\eqn{ThreeLoopFourPtIteration} because of a cancellation between
the last two terms, so they cannot be determined by our four-point
computation. The reason we introduce them is
to handle the subsequent generalization to the $n$-point MHV amplitudes.

\subsection{An ansatz for planar MHV amplitudes to all loop orders}

The resummation and exponentiation of IR singularities 
described by Magnea and Sterman~\cite{MagneaSterman}, 
and the connection to $n$-point amplitudes discussed by Sterman and
Tejeda-Yeomans~\cite{StermanTY}
(both of which we shall review shortly), 
together with the two- and
three-loop iteration formulae, motivate us to propose a 
compact exponentiation of the planar MHV $n$-point amplitudes 
in \SYM\ at $L$ loops.  We propose that
\begin{equation}
{\cal M}_n \equiv 1 + \sum_{L=1}^\infty a^L M_n^{(L)}(\e) 
= \exp\Biggl[\sum_{l=1}^\infty a^l 
          \Bigl(f^{(l)}(\e) M_n^{(1)}(l \e) + C^{(l)} 
               + E_n^{(l)}(\e)  \Bigr) \Biggr] \,.
\label{ExponentialResum}
\end{equation}
In this expression, the factor,
\be
a \equiv { N_c \alpha_s \over 2\pi } (4\pi e^{-\gamma})^\e\,,
\label{alphaberdef}
\ee
keeps track of the loop order of perturbation theory, and coincides 
with the prefactor in brackets in \eqn{LeadingColorDecomposition}.
The quantity $M_n^{(1)}(l\e)$ is the all-orders-in-$\e$ one-loop
amplitude, with the tree amplitude scaled out according 
to~\eqn{RescaledLoopAmplitude}, and with the substitution
$\e \to l\e$ performed.
Each $f^{(l)}(\e)$ is a three-term series in $\e$, 
beginning at $\Ord(\e^0)$,
\be
 f^{(l)}(\e) = f_0^{(l)} + \e f_1^{(l)} + \e^2 f_2^{(l)} \,.
\label{flexp}
\ee
The constants $f_k^{(l)}$ and $C^{(l)}$ are independent of the
number of legs $n$.  They are polynomials in the Riemann values
$\zeta_m$ with rational coefficients, and a uniform degree 
of transcendentality, which is equal to $2l-2+k$ for $f_k^{(l)}$,
and $2l$ for $C^{(l)}$.  
The $f_k^{(l)}$ and $C^{(l)}$ are to be determined by matching
to explicit computations. 
The $E_n^{(l)}(\e)$ are non-iterating $\Ord(\e)$ contributions 
to the $l$-loop amplitudes, which vanish as $\e \rightarrow 0$, 
$E_n^{(l)}(0) = 0$.

Let us first see how \eqn{ExponentialResum} is consistent with the
results up to three loops discussed earlier in this section,
by matching the left- and right-hand sides of the equations
order-by-order in $a$.   The one-loop case is very simple,
since we only have to expand the right-hand side of \eqn{ExponentialResum} 
to $\Ord(a)$.  It agrees with the left-hand side provided that
\begin{equation}
f^{(1)}(\e) = 1\,, \hskip 2 cm C^{(1)} = 0\,, 
\hskip 2 cm E_n^{(1)}(\e) = 0\,.
\label{OneLoopfCE}
\end{equation}
That is, by definition we have absorbed the all-orders-in-$\e$ 
one-loop amplitude into $M_n^{(1)}(\e)$.
(It is possible that for $n>4$ a nonzero value of $E_n^{(1)}(\e)$
could be more natural, given what is known about the structure of
the one-loop amplitudes at $\Ord(\e)$~\cite{DimShift}.)

Next we expand \eqn{ExponentialResum} to two loops, or $\Ord(a^2)$.
Using the one-loop result~(\ref{OneLoopfCE}) to rewrite the
$\Ord(a)$ term in the exponential on the right-hand side
of \eqn{ExponentialResum} as $M_n^{\oneloop}(\e)$, we find that
\begin{equation}
M_n^{\twoloop}(\e) 
= {1 \over 2} \Bigl[ M_n^{\oneloop}(\e) \Bigr]^2
 + f^\twoloop(\e) \, M_n^{\oneloop}(2\e) + C^{(2)}
 + E_n^\twoloop(\e),
\label{TwoLoopOneLoopAgainn}
\end{equation}
which is just the generalization of \eqn{TwoLoopOneLoop} to
$n$ external legs.  Evidence based on collinear limits in favor of
this $n$-leg version, which we shall review in \sect{CollBehaviorSection},
was presented in ref.~\cite{Iterate2};
the values of $f^\twoloop(\e)$ and $C^\twoloop$ given
in \eqns{f2def}{C2def} are independent of $n$.

At the three-loop level, we also use the two-loop 
result~(\ref{TwoLoopOneLoopAgainn}) to rewrite the
$\Ord(a^2)$ term in the exponential on the right-hand side
of \eqn{ExponentialResum} as 
$M_n^\twoloop(\e) - {1\over2} \bigl[ M_n^{\oneloop}(\e) \bigr]^2$.
Matching both sides at $\Ord(a^3)$ gives,
\begin{eqnarray}
M_n^\threeloop(\e) &=& {1\over 6} \Bigl[M_n^\oneloop(\e)\Bigr]^3
            + M_n^\oneloop(\e)\, 
   \biggl\{ M_n^\twoloop(\e) 
          - {1\over 2} \Bigl[M_n^\oneloop(\e)\Bigr]^2 \biggr\}
 \nonumber \\
&&\null
            + f^\threeloop(\e) \, M_n^\oneloop (3\,\e) + C^{(3)}
            + E_n^\threeloop(\e)
 \nonumber \\
 &=& - {1\over 3} \Bigl[M_n^\oneloop(\e)\Bigr]^3
            + M_n^\oneloop(\e)\, M_n^\twoloop(\e)
            + f^\threeloop(\e) \, M_n^\oneloop (3\,\e) + C^{(3)}
            + E_n^\threeloop(\e) \,.\ \ \
\label{ThreeLoopNPtIteration}
\end{eqnarray}
For $n=4$, this equation is equivalent to \eqn{ThreeLoopFourPtIteration},
with the identifications~(\ref{f3def}) and~(\ref{C3def})
for $f^{(3)}(\e)$ and $C^{(3)}$. 

\Eqns{TwoLoopOneLoopAgainn}{ThreeLoopNPtIteration} are special
cases, for $L=2$ and $3$, of a more general $L$-loop 
iteration formula implied by \eqn{ExponentialResum},
\begin{equation}
M_n^{(L)} = X_n^{(L)}\bigl[M_n^{(l)}(\e)\bigr] 
  + f^{(L)}(\e) M_n^{(1)}(L \e) + C^{(L)} 
               + E_n^{(L)}(\e) \,.
\label{iterX}
\end{equation}
The quantities $X_n^{(L)} = X_n^{(L)}[M_n^{(l)}]$ 
only depend on the lower-loop amplitudes $M_n^{(l)}(\e)$ 
with $l<L$.  For $L=2,3$, the values of $X_n^{(L)}$ are,
from \eqns{TwoLoopOneLoopAgainn}{ThreeLoopNPtIteration},
\begin{eqnarray}
X_n^{(2)}\bigl[M_n^{(l)}(\e)\bigr] 
                  &=& {1\over2} \Bigl[ M_n^{(1)} \Bigr]^2  \,, 
\label{X2} \\
X_n^{(3)}\bigl[M_n^{(l)}(\e)\bigr]  &=& - {1\over3} \Bigl[ M_n^{(1)} \Bigr]^3 
            + M_n^{(1)} M_n^{(2)}\,. 
\label{X3}
\end{eqnarray}

Now we establish \eqn{iterX} for arbitrary values of $L$, and provide
a convenient way to compute the functional $X_n^{(L)}[M_n^{(l)}]$.
Taking \eqn{iterX} as a definition of $X_n^{(L)}$, we see that
the full amplitude ${\cal M}_n$ in \eqn{ExponentialResum} can 
also be written as,
\begin{equation}
{\cal M}_n = 1 + \sum_{l=1}^\infty a^l M_n^{(l)}
= \exp\Biggl[
 \sum_{L=1}^\infty a^L ( M_n^{(L)} - X_n^{(L)} ) \Biggr] \,.
\label{Expon2}
\end{equation}
We need to show that the $X_n^{(L)}$ only depend on the 
lower-loop amplitudes $M_n^{(l)}$ with $l<L$.  This result can 
be established inductively on $L$ by comparing the $\Ord(a^L)$ 
terms in the two Taylor expansions.  The coefficient of $a^L$ on the 
left-hand side of \eqn{Expon2} is $M_n^{(L)}$.  On the right-hand side,
$M_n^{(L)}$ occurs explicitly in the $a^L$ term, and with the right
coefficient to match the left-hand side.  Every other term on
the right-hand side depends only on $M_n^{(l)}$ with $l<L$
(using induction for those $X_n^{(L')}$ with $L' < L$).
But $X_n^{(L)}$ must cancel all these other terms for the two Taylor
expansions to agree; hence it also depends only on $M_n^{(l)}$
with $l<L$.

To solve \eqn{Expon2} for $X_n^{(L)}$, we take the logarithm of
both sides, and look at the $L^{\rm th}$ term in the Taylor
expansion of that expression.  We obtain,
\begin{equation}
X_n^{(L)}\bigl[ M_n^{(l)} \bigr]
= M_n^{(L)} 
- \ln\Biggl( 1 + \sum_{l=1}^\infty a^l M_n^{(l)} \Biggr) 
\Biggr\vert_{a^L\ {\rm term}} \,.
\label{Xsol}
\end{equation}
\Eqns{iterX}{Xsol} are key equations; together they provide 
an explicit recipe for writing the $L$-loop amplitude in 
terms of lower-loop amplitudes, plus constant remainders.

From \eqn{Xsol} we can easily recover \eqns{X2}{X3}, as well as
obtain, for example, the next two values of $X_n^{(L)}$:
\begin{eqnarray}
X_n^{(4)}\bigl[M_n^{(l)}(\e)\bigr]  &=& {1\over4} \Bigl[ M_n^{(1)} \Bigr]^4 
            - \Bigl[ M_n^{(1)} \Bigr]^2 M_n^{(2)}
            + M_n^{(1)} M_n^{(3)} 
            + {1\over2} \Bigl[ M_n^{(2)} \Bigr]^2 \,, \\
\label{X4}
X_n^{(5)}\bigl[M_n^{(l)}(\e)\bigr]  &=& - {1\over5} \Bigl[ M_n^{(1)} \Bigr]^5 
            + \Bigl[ M_n^{(1)} \Bigr]^3 M_n^{(2)}
            - \Bigl[ M_n^{(1)} \Bigr]^2 M_n^{(3)}
\nonumber\\ && \hskip0.1cm \null
            - M_n^{(1)} \Bigl[ M_n^{(2)} \Bigr]^2
            + M_n^{(1)} M_n^{(4)} 
            + M_n^{(2)} M_n^{(3)} \,. 
\label{X5}
\end{eqnarray}

Note from \eqn{ExponentialResum} that $f^{(l)}(\e)$ appears multiplied
by $M_n^{(1)}(l\e)$, which has poles beginning only at order $1/\e^2$.
Hence we can absorb any $\Ord(\e^3)$ and higher terms in $f^{(l)}(\e)$
into the definition of the non-iterating contributions $E_n^{(l)}(\e)$.
However, the $\Ord(\e^2)$ terms in $f^{(l)}(\e)$, namely $f_2^{(l)}$, 
cannot be removed because $C^{(l)}$ is asserted to be independent of $n$.
This statement can only be true for one choice of $f_2^{(l)}$;
shifting that value induces a shift proportional
to $n$ in $C^{(l)}$, because $M_n^{(1)}(l\e) \propto n/\e^2$.
The value of $f_2^{(l)}$ can be determined by computing an
$l$-loop amplitude with $n>4$, or else the $l$-loop splitting amplitude
(which may be simpler), as reviewed in \sect{CollBehaviorSection}.

\subsection{Infrared consistency of ansatz}
\label{IRConsistencySection}

In this subsection we discuss the consistency of the exponentiated
$L$-loop ansatz~(\ref{ExponentialResum}) with the resummation
and exponentiation of IR divergences~\cite{Sudakov}, following
the analysis of Magnea and Sterman~\cite{MagneaSterman}, and of
Sterman and Tejeda-Yeomans~\cite{StermanTY}.

A general $n$-point scattering amplitude can be factorized into the 
following form,
\begin{equation}
{\cal M}_n = J\left({Q^2\over\mu^2}, \alpha_s(\mu),\e \right)
           \times S\left( k_i, {Q^2\over\mu^2}, \alpha_s(\mu),\e \right)
           \times h_n\left( k_i,{Q^2\over\mu^2}, \alpha_s(\mu), \e \right) \,,
\label{IRgeneral}
\end{equation}
where $J$ is a jet function, $S$ a soft function, and $h_n$ a hard
remainder function which is finite as $\e \to 0$.
Also, $\mu$ is the renormalization scale,
and $Q$ some physical scale associated with the scattering process
for external momenta $k_i$.

\begin{figure}[t]
\centerline{\epsfxsize 2.3 truein \epsfbox{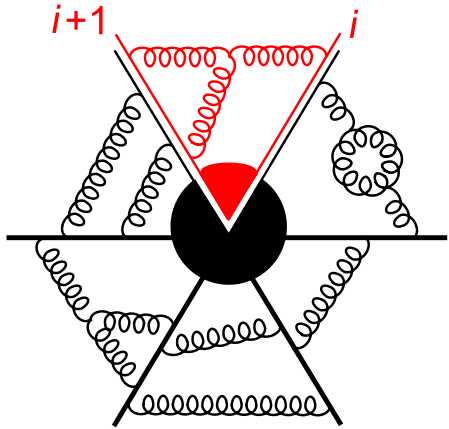}}
\caption{Infrared structure of leading-color scattering amplitudes
for particles in the adjoint representation. 
The straight lines represent hard external states, while the curly
lines carry soft or collinear virtual momenta.  At leading color,
soft exchanges are confined to wedges between the hard lines.}
\label{planarexpFigure}
\end{figure}

Both ${\cal M}_n$ and $h_n$ are vectors in a space of possible color
structures for the process, and $S$ is a matrix.  However,
we shall work in the leading-color (planar) approximation, in
which there is no mixing between the different (color-ordered) color
structures.  Hence $S$ is proportional to the identity matrix.
As pointed out in ref.~\cite{StermanTY}, $S$ is only defined up
to a multiple of the identity matrix, so we can absorb it into
the jet function $J$ at leading color.  \Fig{planarexpFigure}
illustrates that, at leading color, soft exchanges are confined
to wedges between color-adjacent external lines, for example the
lines $i$ and $i+1$ in the figure.  We also consider adjoint 
external states, such as gluons.  By examining the case $n=2$,
it can be seen that the wedge that is being removed
in the figure represents half of the IR singularities
of a Sudakov form factor~\cite{Sudakov}; that is, 
a color-singlet object (such as a Higgs boson) decaying into 
two gluons.  We denote this matrix
element as ${\cal M}^{[gg\to1]}(s_{i,i+1}/\mu^2,\alpha_s(\mu),\e)$.

Because \SYM\ is conformally invariant (the $\beta$ function vanishes),
$\alpha_s$ may be set to a constant everywhere.  Thus the leading-color
IR structure of $n$-point amplitudes in \SYM\ may be rewritten as,
\begin{equation}
{\cal M}_n = \prod_{i=1}^n 
 \Biggl[  {\cal M}^{[gg\to1]}\left( {s_{i,i+1}\over\mu^2},\alpha_s,\e \right) 
 \Biggr]^{1/2}
 \times h_n\left( k_i, \mu, \alpha_s, \e \right) \,,
\label{IRplanar}
\end{equation}
where $h_n$ is no longer a color-space vector.

%

For a general theory, the Sudakov form factor at scale $Q^2$ 
can be written as~\cite{MagneaSterman}
\begin{eqnarray}
{\cal M}^{[gg\to1]}\biggl( {Q^2 \over \mu^2} , \alpha_s(\mu), \e \biggr)
&=& \exp \Biggl\{ {1\over2}
  \int_0^{-Q^2} {d \xi^2 \over \xi^2} \biggl[ 
                  {\cal K}^{[g]}\bigl(\alpha_s(\mu),\e \bigr)
        + {\cal G}^{[g]}\biggl( -1, \bar\alpha_s\Bigl( {\mu^2\over\xi^2}, 
                                   \alpha_s(\mu),\e \Bigr), \e \biggr)
\nonumber\\
&&\hskip1.0cm
        + {1\over2} \int_{\xi^2}^{\mu^2} {d \tilde \mu^2 \over \tilde \mu^2} 
               \gamma_K^{[g]}\biggl( 
                              \bar\alpha_s\Bigl( {\mu^2\over\tilde \mu^2}, 
                                   \alpha_s(\mu),\e \Bigr) \biggr)
 \biggr] \Biggr\} \,,
\label{SudakovFF}
\end{eqnarray}
where $\gamma_K^{[g]}$ denotes the soft or (Wilson line) cusp 
anomalous dimension, which will produce a $1/\e^2$ pole after 
integration.  The function ${\cal K}^{[g]}$ is a series
of counterterms (pure poles in $\e$), while ${\cal G}^{[g]}$
includes non-singular dependence on $\e$ before integration,
and produces a $1/\e$ pole after integration.

In \SYM, $\alpha_s(\mu)$ is a constant, and the running coupling
$\bar\alpha_s(\mu^2/\tilde\mu^2,\alpha_s,\e)$ in $4-2\e$ dimensions
has only trivial (engineering) dependence on the scale,
\be
\bar\alpha_s\Bigl( {\mu^2\over\tilde \mu^2},\alpha_s(\mu), \e \Bigr)
= \alpha_s \times \biggl( {\mu^2\over \tilde\mu^2} \biggr)^\e 
\Bigl( 4 \pi e^{-\gamma} \Bigr)^\e \,.
\label{alphabar}
\ee
This simple dependence makes it very easy to perform the integrals over
$\xi$ and $\tilde\mu$.

Following refs.~\cite{MagneaSterman,StermanTY}, we expand ${\cal
K}^{[g]}$, $\gamma_K^{[g]}$, and ${\cal G}^{[g]}$ in powers of
$\alpha_s$,
\begin{eqnarray}
{\cal K}^{[g]}(\alpha_s,\e) &=& 
\sum_{l=1}^\infty 
{1 \over 2 l \e }
a^l \,
 \hat\gamma_K^{(l)}
 \,, \label{Kexpand} \\
\gamma_K^{[g]}\biggl( 
  \bar\alpha_s \Bigl( {\mu^2\over\tilde \mu^2},\alpha_s, \e \Bigr)
 \biggr) 
&=& \sum_{l=1}^\infty  
a^l
\Bigl( {\mu^2 \over \tilde \mu^2} \Bigr)^{l\e} \hat\gamma_K^{(l)} \,,
\label{gammaKexpand} \\
{\cal G}^{[g]}\biggl( -1, \bar\alpha_s\Bigl( {\mu^2\over\xi^2}, 
                              \alpha_s, \e \Bigr), \e \biggr)
 &=&  \sum_{l=1}^\infty 
a^l
\Bigl({\mu^2 \over  \xi^2} \Bigr)^{l\e} {\hat{\cal G}}_0^{(l)} \,,
\label{Gexpand}
\end{eqnarray}
where $a$ is defined in \eqn{alphaberdef} and the hats are a reminder
that the leading-$N_c$ dependence has also been removed in
eqs.~(\ref{Kexpand}), (\ref{gammaKexpand}) and (\ref{Gexpand}).
That is, the perturbative coefficients 
(defined with expansion parameter $\alpha_s/(2\pi)$) 
have a leading-color dependence on $N_c$ of,
\begin{equation}
\gamma_K^{(l)} = \hat\gamma_K^{(l)} \, N_c^l \,, \hskip 2 cm 
{\cal G}_0^{(l)} = \hat {\cal G}_0^{(l)} \, N_c^l \,.
\label{NcRescaledGammaG}
\end{equation}
We can suppress the $[g]$ label because the $N=4$ MHV
amplitudes are all related by supersymmetry Ward identities~\cite{SWI}, 
so that the corresponding functions for external gluinos, {\it etc.}, 
are the same as for gluons.
\Eqn{Kexpand} follows from solving eqs.~(2.12) and (2.13) of 
ref.~\cite{MagneaSterman} in the conformal case ($\beta \equiv 0$).
In this case, ${\cal K}^{[g]}$ contains only single poles in $\e$,
which are simply related to $\gamma_K^{[g]}$.

The integral over ${\cal G}$ is very simple,
\begin{equation}
\int_0^{-Q^2} {d \xi^2 \over \xi^2} {\cal G}^{[g]} =
- \sum_{l=1}^\infty\ {a^l\over l\e} \,
   \Bigl( {\mu^2 \over -Q^2} \Bigr)^{l\e}   {\hat{\cal G}}_0^{(l)} 
\,.
\label{IntG}
\end{equation}
The first integral over $\gamma_K$ gives,
\begin{equation} 
\int_{\xi^2}^{\mu^2} {d \tilde \mu^2 \over \tilde \mu^2} 
               \gamma_K^{[g]}  
= \sum_{l=1}^\infty \ { a^l\over l \e}  \,
  \biggl[  \Bigl( {\mu^2 \over \xi^2} \Bigr)^{l\e} - 1 \biggr]
  \hat\gamma_K^{(l)} 
\,.
\label{IntgammaK1}
\end{equation}

Adding the ${\cal K}^{[g]}$ term to $1/2$ of \eqn{IntgammaK1},
using \eqn{Kexpand}, we see that the ``$-1$'' is cancelled.
Then the integral over $\xi$ is properly regulated, and 
evaluates to
\be
- {1\over2} \sum_{l=1}^\infty { a^l \over (l\e)^2 }
 \Bigl( {\mu^2 \over -Q^2} \Bigr)^{l\e} 
  \hat\gamma_K^{(l)} \,.
\label{IntgammaK2}
\ee
Combining this result with \eqn{IntG} gives
\begin{equation}
{\cal M}^{[gg\to1]}\biggl( {Q^2 \over \mu^2} , \alpha_s(\mu), \e \biggr)
= \exp \Biggl[ - {1\over4} \sum_{l=1}^\infty
  a^l \Bigl( {\mu^2 \over -Q^2} \Bigr)^{l\e} 
 \biggl( { \hat\gamma_K^{(l)} \over (l\e)^2 }
       + { 2 \hat{\cal G}_0^{(l)} \over l\e } \biggr) \Biggr] \,.
\label{Mfinal}
\end{equation}

We need \eqn{Mfinal} for a neighboring pair of legs $i,i+1$
in the $n$-point amplitude, so that $Q^2$ should be replaced by 
the invariant $s_{i,i+1}$.  Taking the product over all $i$,
\eqn{IRplanar} becomes
\begin{eqnarray}
{\cal M}_n &=& \exp\Biggl[ - {1\over8} \sum_{l=1}^\infty
  a^l \, 
 \Bigl( \hat\gamma_K^{(l)} +  2 l \hat{\cal G}_0^{(l)} \e \Bigr) 
  { 1 \over (l\e)^2 }
\sum_{i=1}^n   \Bigl( {\mu^2 \over -s_{i,i+1}} \Bigr)^{l\e} \Biggr]
 \times h_n \,.
\label{IRplanarNew1}
\end{eqnarray}
We may rearrange this a bit, to give
\begin{eqnarray}
{\cal M}_n
&=& \exp\Biggl[ \sum_{l=1}^\infty  a^l \, 
   f^{(l)}(\e) \hat{I}_n^{(1)}(l\e) \Biggr]
 \times \tilde{h}_n \,,
\label{IRplanarNew2}
\end{eqnarray}
where $f^{(l)}(\e)$ is defined in \eqn{flexp},
with the identifications,
\begin{eqnarray}
f^{(l)}_0 &=& {1\over4} \, \hat\gamma_K^{(l)} \,, 
\label{f0toK}\\
f^{(l)}_1 &=& {l \over 2} \, \hat{\cal G}_0^{(l)} \,,
\label{f1toG}
\end{eqnarray}
and
\be
\hat{I}_n^{(1)}(\e) 
= - {1\over2} {1\over\e^2} \sum_{i=1}^n 
 \Bigl( {\mu^2 \over -s_{i,i+1}} \Bigr)^{\e} \,.
\label{Ihatn}
\ee
Here $\tilde{h}_n$ differs from $h_n$ by a finite shift, due to the
$\Ord(\e^2)$ terms in $f^{(l)}(\e)$, which we introduce to help make
contact with the exponentiated ansatz~(\ref{ExponentialResum}).  Using
$f^{(1)}(\e) = 1$ and \eqns{f2def}{f3def}, we may read off the first
few loop orders,
\begin{eqnarray}
\hat \gamma_K^{(1)} &=& 4 \,, \nn\\
\hat \gamma_K^{(2)} &=& - 4 \zeta_2 \,, \label{GammaValues} \\
\hat \gamma_K^{(3)} &=& 22 \,\zeta_4 \,, \nn
\end{eqnarray}
and
\begin{eqnarray}
\hat{\cal G}_0^{(1)} &=& 0 \,, \nn\\
\hat{\cal G}_0^{(2)} &=& - \zeta_3 \,, \label{calGValues}\\
\hat{\cal G}_0^{(3)} &=& 4 \zeta_5 + {10\over 3} \zeta_2 \zeta_3 \,. \nn
\end{eqnarray}

The quantity $\hat{I}_n^{(1)}(\e)$ is a function that
captures the divergences of the planar one-loop $n$-point 
amplitudes in \SYM~\cite{IROneLoop},
after extracting the leading-$N_c$ dependence as in
ref.~\cite{TwoLoopSplitting}.  The $\hat{I}_n^{(1)}$
defined in eq.~(8.9) of ref.~\cite{TwoLoopSplitting}
contained a prefactor of $e^{-\e\psi(1)}/\Gamma(1-\e)$,
following conventions of Catani~\cite{CataniIR}.
Here we adopt a convention closer to that of Sterman
and Tejeda-Yeomans~\cite{StermanTY}, without such a prefactor.
The difference between eq.~(8.9) of ref.~\cite{TwoLoopSplitting} 
and \eqn{Ihatn} above is a finite quantity,
because $e^{-\e\psi(1)}/\Gamma(1-\e) = 1 + \Ord(\e^2)$.
Finite remainders will differ between the two conventions.

Starting from \eqn{IRplanarNew2}, and using the fact that the
difference between $M_n^{(1)}(l\e)$ and $\hat{I}_n^{(1)}(l\e)$
is finite, we can reshuffle the finite terms once more to obtain,
\begin{equation}
{\cal M}_n = \exp\Biggl[ \sum_{l=1}^\infty
        a^l \, 
   \Bigl( f^{(l)}(\e) M_n^{(1)}(l\e) + h_n^{(l)}(\e) \Bigr) \Biggr] \,.
\label{IRplanarNew3}
\end{equation}
We have moved the hard function into the exponent without loss
of generality, because we allow for a new function $h_n^{(l)}$ at
each order $l$.

Finally we compare the exponentiated ansatz~(\ref{ExponentialResum})
with the exponentiation of the IR divergences~(\ref{IRplanarNew3}).
We see that they agree if we identify
\be
h_n^{(l)}(k_i,\e) = C^{(l)} + E_n^{(l)}(\e) \,.
\label{hCErelation}
\ee
In some sense, the content of the iterative structure of 
planar \SYM, beyond the level of consistency with IR 
resummation, is that the (suitably-defined) exponentiated
hard remainders $h_n^{(l)}(k_i,\e)$ approach constants, 
independent of the kinematics and of $n$, as $\e \to 0$, since
$E_n^{(l)}(\e)$ is of $\Ord(\e)$.

\subsection{Finite Remainders}

Next we shall obtain iterative formulae for two series of functions:
the $\hat I_n^{(L)}$ governing IR divergences for the $L$-loop $n$-point
planar amplitudes, and the $F_n^{(L)}$ representing the finite remainders
of the amplitudes, after subtracting these divergences.
The formulae will be very similar in form to the full amplitude
relation~(\ref{iterX}).

Following the structure uncovered explicitly at one, two, and three 
loops~\cite{IROneLoop,CataniIR,StermanTY},
we define the finite remainder for the $L$-loop amplitude by writing
\begin{equation}
M_n^{(L)}(\e) = \sum_{l=0}^{L-1} \hat I_n^{(L-l)}(\e) \, M_n^{(l)}(\e)
 + F_n^{(L)}(\e) \,,
\label{FnLdef0}
\end{equation}
or
\begin{equation}
F_n^{(L)}(\e) = M_n^{(L)} - \sum_{l=0}^{L-1} \hat I_n^{(L-l)} \, M_n^{(l)} \,,
\label{FnLdef}
\end{equation}
where $M_n^{(0)} \equiv 1$.
We insert the iteration formula~(\ref{iterX}) for the first term,
$M_n^{(L)}$, on the right-hand side of \eqn{FnLdef}.
We split $M_n^{(1)}(L \e) \to \hat I_n^{(1)}(L \e) + F_n^{(1)}(L \e)$
in this formula.
For the lower-loop amplitudes, $M_n^{(l)}$, we recursively substitute
in the finite-remainder formula for smaller values of $l$,
\begin{equation}
M_n^{(l)} = \sum_{k=0}^{l-1} \hat I_n^{(l-k)} M_n^{(k)} + F_n^{(l)} \,.
\label{lowerloopfinite}
\end{equation}
At this point, the expression for $F_n^{(L)}(\e)$
is a polynomial in $\hat I_n^{(l)}$ and $F_n^{(l)}$,
which has the special property that there are no mixed $\hat{I}$-$F$ terms.
(If there had been such terms, it would have signaled an inconsistency.) 
We can remove the pure-$\hat{I}$ terms by choosing $\hat I_n^{(L)}$ 
to cancel them.
The resulting finite expression gives $F_n^{(L)}(\e)$ as a polynomial
in the lower-loop $F_n^{(l)}(\e)$.  

We find that the solutions for $\hat I_n^{(L)}(\e)$ and $F_n^{(L)}(\e)$ 
are expressible in terms of the same 
function $X_n^{(L)}$ defined in \eqn{Xsol},
but where the role of $M_n$ is played instead by $-\hat I_n$ and $F_n$,
respectively:
\begin{eqnarray}
\hat I_n^{(L)}(\e) &=& - X_n^{(L)}\bigl[ - \hat I_n^{(l)}(\e) ]
                  + f^{(L)}(\e) \hat I_n^{(1)}(L \e) \,, 
\label{Isol} \\
F_n^{(L)}(\e) &=& X_n^{(L)}\bigl[ F_n^{(l)}(\e) ]
                  + f^{(L)}(\e) F_n^{(1)}(L \e) + C^{(L)} + E_n^{(L)}(\e) \,. 
\label{Fsol}
\end{eqnarray}
The Taylor expansion~(\ref{Xsol}) can be used to evaluate \eqns{Isol}{Fsol}
to any desired loop order.

Because the form of \eqn{Fsol} for $F_n^{(L)}(\e)$ is completely analogous to 
the iteration formula~(\ref{iterX}) for the full amplitude $M_n^{(L)}(\e)$,
we see that the finite remainders can be exponentiated as,
\begin{equation}
{\cal F}_n(\e) \equiv 1 + \sum_{L=1}^\infty a^L F_n^{(L)}(\e)
=  \exp\Biggl[\sum_{l=1}^\infty a^l 
          \Bigl(f^{(l)}(\e) F_n^{(1)}(l \e) + C^{(l)} 
               + E_n^{(l)}(\e)  \Bigr) \Biggr] \,.
\label{FepsResum}
\end{equation}
Letting $\e\to0$, we have
\begin{equation}
{\cal F}_n(0) \equiv 1 + \sum_{L=1}^\infty a^L F_n^{(L)}(0)
=  \exp\Biggl[\sum_{l=1}^\infty a^l 
          \Bigl(f^{(l)}_0 F_n^{(1)}(0) + C^{(l)} \Bigr) \Biggr] \,,
\label{F0Resum}
\end{equation}
where
\be
f^{(l)}_0  \equiv f^{(l)}(0) \,.
\ee
Using \eqn{f0toK} we may then rewrite this as,
\begin{equation}
{\cal F}_n(0) = 
\exp\Biggl[{1\over 4} \gamma_K F_n^{(1)}(0) + C  \Biggr] \,.
\label{F0ResumFinal}
\end{equation}
The soft anomalous dimension is
\begin{eqnarray}
\gamma_K &=& \sum_{l=1}^\infty \hat\gamma_K^{(l)} a^l
 =  4\, a  - 4 \zeta_2 \,a^2 +  22 \zeta_4 \,a^3 + \cdots \,, 
\label{gammaKa}
\end{eqnarray}
where we used \eqn{GammaValues}.  Similarly,
from eqs.~(\ref{C2def}), (\ref{C3def}) and~(\ref{OneLoopfCE}), we have,
\begin{eqnarray}
C &=& \sum_{l=1}^\infty C^{(l)} a^l = 
 - {1\over 2} \zeta_2^2 \, a^2 +
\Biggl[
   \biggl( {341\over 216} \, + {2\over 9} c_1 \biggl) \zeta_6
 + \biggl( - {17\over 9} + {2\over 9} c_2 \biggr)\zeta_3^2 \Biggr]\, a^3 
 + \cdots \,.
\label{Ca}
\end{eqnarray}
As mentioned below \eqn{C3def}, the rational numbers $c_1$ and $c_2$ are yet
to be determined.  The resummation~(\ref{F0ResumFinal}) 
of the finite remainders of the MHV amplitudes, 
as a consequence of the exponentiated ansatz~(\ref{ExponentialResum}), 
is one of the key results of this paper.  

For the $F_n^{(l)}(0)$, the argument $l\e$ 
in \eqn{Fsol} has disappeared as $\e\to0$.  
Hence we can recursively substitute back to obtain formulas
solely in terms of $F_n^{(1)}(0)$.  Equivalently, we can series expand
the exponential in \eqn{F0Resum} or (\ref{F0ResumFinal}).  The first
few values are,
\begin{eqnarray}
F_n^{(2)}(0) &=& {1\over 2} \Bigl(F_n^{(1)}(0)\Bigr)^2 +
       f^{(2)}_0 F_n^{(1)}(0) + C^{(2)} \,,
\label{F2iter} \\
F_n^{(3)}(0) &=& - {1\over 3} \Bigl(F_n^{(1)}(0)\Bigr)^3 +
       F_n^{(1)}(0) F_n^{(2)}(0)  +
       f^{(3)}_0 F_n^{(1)}(0) + C^{(3)} \\
&=& {1\over 6} \Bigl(F_n^{(1)}(0)\Bigr)^3 +
       f^{(2)}_0 \Bigl(F_n^{(1)}(0)\Bigr)^2 +
       \Bigl(f^{(3)}_0 +  C^{(2)}\Bigr) F_n^{(1)}(0)
       + C^{(3)} \,, 
\label{F3iter} \\
F_n^{(4)}(0) &=& {1\over4} \Bigl( F_n^{(1)}(0) \Bigr)^4 
            - \Bigl( F_n^{(1)}(0) \Bigr)^2 F_n^{(2)}(0)
            + F_n^{(1)}(0) F_n^{(3)}(0) 
            + {1\over2} \Bigl( F_n^{(2)}(0) \Bigr)^2 
\nonumber \\ && \hskip 0.05cm
            + f^{(4)}_0 F_n^{(1)}(0) + C^{(4)} \\
&=& {1\over 24} \Bigl( F_n^{(1)}(0) \Bigr)^4 
  + {1\over2} f^{(2)}_0 \Bigl( F_n^{(1)}(0) \Bigr)^3 
  + {1\over2} \Bigl( [ f^{(2)}_0 ]^2 + 2 f^{(3)}_0 + C^{(2)} \Bigr) 
     \Bigl( F_n^{(1)}(0)\Bigr)^2
\nonumber \\ && \hskip 0.05cm
  + \Bigl( f^{(4)}_0 + f^{(2)}_0 C^{(2)} + C^{(3)} \Bigr) 
        F_n^{(1)}(0)
   + C^{(4)} + {1\over2} [ C^{(2)} ]^2 \,.
\label{F4iter}
\end{eqnarray}
Thus, starting from the ansatz~(\ref{ExponentialResum}),
we have succeeded in expressing the $n$-point $L$-loop MHV finite
remainders directly in terms of the one-loop finite remainders.

We remark that the two-loop result~(\ref{F2iter})  
differs slightly (in the constant term) from the corresponding 
eq.~(16) for $n=4$ in ref.~\cite{Iterate2}.  The reason is that
the definition of the two-loop divergence used there, interpreted
in terms of $\hat{I}_n^{(2)}$, does not obey \eqn{Isol} for $L=2$,
but differs from that $\hat{I}_n^{(2)}$ by a finite ($\Ord(\e^0)$)
amount.  The definition we use here is more convenient because 
of its simple generalization to higher loops.

The one-loop finite remainders $F_n^{(1)}(0)$ for the MHV amplitudes
in \SYM\ were evaluated for all $n$ in ref.~\cite{NeqFourOneLoop}, 
using the unitarity method. 
Modifying those results to the conventions of this paper,
the finite terms are explicitly, for all $n\geq5$,
\begin{equation}
F_n^{(1)}(0) = {1 \over 2} \sum_{i=1}^n g_{n,i} \,,
\label{OneLoopFiniteRemainder}
\end{equation}
where
\begin{eqnarray}
g_{n,i} &=&
-\sum_{r=2}^{\lfloor n/2 \rfloor -1}
  \ln \Biggl({ -\tn{r}{i}\over -\tn{r+1}{i} }\Biggr)
  \ln \Biggl({ -\tn{r}{i+1}\over -\tn{r+1}{i} }\Biggr) +
D_{n,i} + L_{n,i} + {3\over2} \zeta_2 \,,
\label{UniversalFunci}
\end{eqnarray}
and where $\lfloor x \rfloor$ is the greatest integer less than or equal
to $x$.  Here $\tn{r}{i} = (k_i + \cdots + k_{i+r-1})^2$
are the momentum invariants, so that $\tn1i = 0$ and $\tn2i = s_{i,i+1}$.
(All indices are understood to be $\mod n$.)
The form of $D_{n,i}$ and $L_{n,i}$ depends upon whether $n$ is odd or even.
For $n=2m+1$,
\begin{eqnarray}
D_{2m+1,i} &=& -\sum_{r=2}^{m-1}
\li2 \Biggl( 1- { \tn{r}{i} \tn{r+2}{i-1}
\over \tn{r+1}{i} \tn{r+1}{i-1} } \Biggr) \,,
\label{Dodd} \\
L_{2m+1,i} &=& -{ 1\over 2}
  \ln \Biggl({ -\tn{m}{i}\over -\tn{m}{i+m+1}  } \Biggr)
  \ln \Biggl({ -\tn{m}{i+1}\over -\tn{m}{i+m} } \Biggr) \,,
\label{Lodd}
\end{eqnarray}
whereas for $n=2m$,
\begin{eqnarray}
D_{2m,i} &=& -\sum_{r=2}^{m-2}
\li2 \Biggl( 1- { \tn{r}{i} \tn{r+2}{i-1}
\over \tn{r+1}{i} \tn{r+1}{i-1} }  \Biggr)
- {1 \over 2} \li2 \Biggl( 1- { \tn{m-1}{i} \tn{m+1}{i-1}
\over \tn{m}{i} \tn{m}{i-1}} \Biggr) \,,
\label{Deven} \\
L_{2m,i} &=& - {1\over 4}
  \ln \Biggl({ -\tn{m}{i}\over -\tn{m}{i+m+1}  } \Biggr)
  \ln \Biggl({ -\tn{m}{i+1}\over -\tn{m}{i+m} } \Biggr) \,.
\label{Leven}
\end{eqnarray}
For $n=4$ the above formula does not hold, but the finite 
remainder is simply,
\begin{equation}
F_4^{(1)}(0) 
= {1\over 2} \ln^2\biggl({-t\over-s}\biggr) + 4 \zeta_2 \,. 
\label{STYF10}
\end{equation}

Assuming that the exponentiated ansatz~(\ref{ExponentialResum}) holds,
then the exponentiated finite remainders ${\cal F}_n(0)$ given 
in~\eqn{F0ResumFinal} are completely determined to all loop orders,
in terms of the one-loop remainders $F_n^{(1)}(0)$ just presented, 
plus the series of constants $\gamma_K$ and $C$ given 
in~\eqns{gammaKa}{Ca}.


\section{Collinear behavior and consistency of all-$n$ ansatz}
\label{CollBehaviorSection}

In this section we discuss the consistency of the $n$-point iterative
ansatz~(\ref{iterX}) with the behavior of amplitudes under
factorization.  In a supersymmetric theory, MHV amplitudes have
no multi-particle poles; the residues vanish by a supersymmetry
Ward identity~\cite{SWI}.  This property is manifest in our ansatz, 
because neither the tree amplitude $A_n^{\tree,\,{\rm MHV}}(1,2,\ldots,n)$ 
nor the one-loop amplitude $M_{n-1}^{(1)}(l\e)$ contain such poles.  
Hence only factorizations as pairs of momenta become collinear 
need to be analyzed. 
 
In general, color-ordered amplitudes $A_n^{(L)}(1, 2, \ldots, n)$ 
satisfy simple
properties as the momenta of two color-adjacent legs $k_i$, $k_{i+1}$
become collinear,~\cite{TreeReview,NeqFourOneLoop,OneLoopSplitting,
KosowerSplit,OneLoopReview},
\begin{eqnarray}
 &&A_n^\Lloop(\ldots,i^{\lambda_i},(i+1)^{\lambda_{i+1}},\ldots)
  \, \longrightarrow
\sum_{l = 0}^L \sum_{\lambda=\pm}
  \Split^\lloop_{-\lambda}(z;i^{\lambda_i}\kern-1pt,(i+1)^{\lambda_{i+1}})
  A_{n-1}^{(L-l)}(\ldots,P^\lambda \kern-5pt,\ldots) \,. \hskip 1 cm
\label{LoopSplit}
\end{eqnarray}
The index $l$ sums over the different loop orders of contributing
splitting amplitudes $\Split_{-\lambda}^\lloop$, while $\lambda$ sums
over the helicities of the intermediate leg $k_P=-(k_i + k_{i+1})$, 
and $z$ is the longitudinal momentum fraction of $k_i$, 
$k_i \approx -z k_P$. 
The splitting amplitudes are universal and gauge invariant.

The tree-level splitting amplitudes $\Split_{-\lambda}^{(0)}$ are
the same in \SYM\ as in QCD. At loop level, the \SYM\ splitting
amplitudes are all proportional to the tree-level ones.  The
proportionality factors depend only on $z$ and $\e$, not on the 
helicity configuration, nor (except for a trivial dimensional factor) 
on kinematic invariants~\cite{NeqFourOneLoop}.  
It is thus convenient to write
the $L$-loop planar splitting amplitudes in terms of
``renormalization'' factors $r_S^{(L)}(\e;z,s)$, defined
by
\begin{eqnarray}
 \Split_{-\lambda_P}^\Lloop(1^{\lambda_1},2^{\lambda_2})
& =&  \,
  r_S^{(L)} (\e;z,s)
      \Split_{-\lambda_P}^\tree(1^{\lambda_1},2^{\lambda_2}) \,,
\label{LLooprDef}
\end{eqnarray}
where $s = (k_1+k_2)^2$.

Using \eqns{LoopSplit}{LLooprDef}, we see that the
amplitude ratios
$M_n^\Lloop(\e) \equiv A_n^\Lloop/A_n^\tree$
behave in collinear limits as,
\begin{eqnarray}
M_n^{\oneloop}(\e) &\rightarrow& M_{n-1}^{\oneloop}(\e)
+ r_S^{\oneloop}(\e) \,,
\label{OneLoopCollinear}\\
M_n^{\twoloop}(\e) & \rightarrow & M_{n-1}^{\twoloop}(\e)
+ r_S^{\oneloop}(\e) M_{n-1}^{\oneloop}(\e) + r_S^{(2)}(\e) \,,
\label{TwoLoopCollinear}
\end{eqnarray}
or at three loops,
\begin{eqnarray}
M_n^{\threeloop}(\e) & \rightarrow & M_{n-1}^{\threeloop}(\e)
+ r_S^{\oneloop}(\e) M_{n-1}^{\twoloop}(\e)
+ r_S^{\twoloop}(\e) M_{n-1}^{\oneloop}(\e)
+ r_S^{\threeloop}(\e) \,,
\label{ThreeLoopCollinear}
\end{eqnarray}
where $r_S^{(0)}(\e) \equiv 1$ and we have suppressed all functional
arguments except for $\e$.

The one-loop splitting amplitudes in \SYM\ have been
calculated to all orders in $\e$, with the result~\cite{OneLoopSplitting},
\begin{eqnarray}
r_S^{\oneloop}(\e;z,s) &=&  { \cgh  \over \e^2}
  \biggl( { \mu^2 \over -s } \biggr)^\e
  \biggl[ - {\pi \e \over \sin(\pi \e) } \biggl( { 1-z \over z } \biggr)^\e
          + 2 \, \sum_{k=0}^\infty \e^{2k+1} \,
                   \Li_{2k+1}\biggl( {- z \over 1-z } \biggr) \biggr] \,,
\label{OneLooprSUSY}
\end{eqnarray}
where $\Li_n$ is the $n$-th polylogarithm (defined in \eqn{PolyLogDef}),
and
\begin{equation}
\cgh = {e^{\e \gamma} \over 2}
{\Gamma(1+\e) \Gamma^2(1-\e) \over \Gamma(1-2\e) } \,.
\label{cGammahatDef}
\end{equation}

In refs.~\cite{Iterate2,TwoLoopSplitting}, the two-loop
splitting amplitudes in \SYM\ were computed through $\Ord(\e^0)$ using the
unitarity method as described in ref.~\cite{KosowerSplit}.  The result
of this computation is a very simple formula, expressing the
two-loop splitting amplitude in terms of the one-loop one,
\begin{equation}
r^{(2)}_S(\e;z, s)  
= {1 \over 2} \Bigl( r^\oneloop_S(\e;z, s) \Bigr)^2
     + f^\twoloop(\e) \, r^\oneloop_S(2\e;z, s) + \Ord(\e) \,,
\label{OneLoopTwoLoopSplit}
\end{equation}
where $f^\twoloop(\e)$ is given in \eqn{f2def}.
(This result was actually obtained before the iterative
relation~(\ref{TwoLoopOneLoop}), and motivated its discovery.) 

The consistency of the $n$-point 
ansatz~(\ref{iterX}) for $L=2$ (with $X_n^{(2)}$
given by \eqn{X2}) may be easily confirmed using
these splitting functions~\cite{Iterate2}.  
Inserting the collinear behavior of the one-loop
amplitudes~(\ref{OneLoopCollinear}) into the
right-hand side, we obtain,
\begin{eqnarray}
M_{n}^\twoloop(\e) &\rightarrow &
 {1 \over 2} \Bigl(M_{n-1}^{\oneloop}(\e) + r_S^{\oneloop}(\e) \Bigr)^2
      + f^\twoloop(\e) \, (M_{n-1}^{\oneloop}(2\e) + r_S^{\oneloop}(2\e)) 
      - {1\over 2} \zeta_2^2  \nn \\
 &=&  M_{n-1}^\twoloop(\e)  + r_S^{\oneloop}(\e) M_{n-1}^\oneloop(\e)
        +  r_S^{\twoloop}(\e) \,,
\label{Check2LoopColl}
\end{eqnarray}
where we used \eqns{iterX}{OneLoopTwoLoopSplit} to
rearrange the expression into the required form
(\ref{TwoLoopCollinear}) for correct two-loop collinear behavior.
Since there are no multi-particle poles in the MHV case, 
\eqn{Check2LoopColl} confirms that the ansatz~(\ref{iterX}) has 
the correct factorization properties in all channels at two loops.  

Similarly, we can require that the ansatz~(\ref{iterX})
is consistent with collinear factorization beyond two loops,
and thereby obtain an iterative ansatz for the planar $L$-loop
splitting amplitudes in \SYM.  For $L=3$, the ansatz reads,
\begin{eqnarray}
M_n^\threeloop(\e) &=& - {1\over 3} \Bigl[M_n^\oneloop(\e)\Bigr]^3
            + M_n^\oneloop(\e)\, M_n^\twoloop(\e)
            + f^\threeloop(\e) \, M_n^\oneloop (3\,\e) + C^{(3)}
            + \Ord(\e) \,.
\label{ThreeLoopNPointIteration}
\end{eqnarray}
If we insert the properties of one- and two-loop
amplitudes~(\ref{OneLoopCollinear}) and~(\ref{TwoLoopCollinear})
into the collinear limit of the right-hand side 
of \eqn{ThreeLoopNPointIteration}, we obtain,
\begin{eqnarray}
M_n^\threeloop(\e) &\rightarrow &
   - {1\over 3} \Bigl[M_{n-1}^\oneloop(\e) + r_S^{\oneloop}(\e)\Bigr]^3 \nn \\
&& \null
            + \Bigl(M_{n-1}^\oneloop(\e)+ r_S^{\oneloop}(\e) \Bigr) \,
             \Bigl(M_{n-1}^{\twoloop}(\e)
    + r_S^{\oneloop}(\e) M_{n-1}^{\oneloop}(\e) 
    + r_S^{\twoloop}(\e)  \Bigr) \nn \\
&& \null
        + f^\threeloop(\e) \,
        \Bigl(M_{n-1}^\oneloop (3\,\e) + r_S^\oneloop(3\,\e) \Bigr) 
        + C^{(3)} + \Ord(\e) \,.
\label{ThreeLoopNPointCollinear}
\end{eqnarray}
After rearranging terms, we can get consistency with 
\eqn{ThreeLoopCollinear}, provided that the three-loop
splitting function obeys,
\begin{equation}
r^{(3)}_S(\e)  
= -{1 \over 3} \Bigl( r^\oneloop_S(\e) \Bigr)^3
     + r^\oneloop_S(\e)  r^\twoloop_S(\e)
     + f^\threeloop(\e) \,  r^\oneloop_S(3\e) + \Ord(\e) \,.
\label{OneLoopThreeLoopSplit}
\end{equation}

By repeating this exercise at $L$ loops, and collecting the
terms that are independent of $M_{n-1}^{(l)}$, we see that
the relation,
\be
r_S^{(L)}(\e)  
= X^{(L)}\bigl[ r^{(l)}_S(\e) \bigr]
  + f^{(L)}(\e) \, r_S^{(L)}(L\e) + \Ord(\e)  \,,
\label{IterateILoopSplit}
\end{equation}
is the one required for consistency.  (We have dropped the subscript $n$
from $X_n^{(L)}$ because it is out of place here, but it is
the same function of lower-loop quantities defined in \eqn{Xsol}.)
In other words, the $L$-loop splitting amplitude functions $r_S^{(L)}$ 
obey exactly the same type of iterative relation as the 
scattering amplitudes $M_n^{(L)}$, but without
the ``inhomogeneous'' constant terms $C^{(L)}$.
Because the one-loop splitting amplitude $r_S^{(1)}(L\e)$ begins 
at order $\Ord(\e^{-2})$,
the relation~(\ref{IterateILoopSplit}) allows the $\Ord(\e^2)$
coefficient of $f^{(L)}(\e)$, namely $f_2^{(L)}$, to be extracted
from the $\Ord(\e^0)$ term in the $L$-loop splitting amplitude.

\section{Anomalous dimensions and Sudakov form factors}
\label{AnomalousDimensionSection}

The soft anomalous dimension $\gamma_K$ controlling the $1/\e^2$
IR singularities of the loop amplitudes arises from an edge of phase
space, the Sudakov region, where a hard line can only emit soft gluons.
In the loop amplitudes these gluons are virtual, of course, but 
they are related to real soft-gluon emission by the cancellation 
of infrared poles in infrared-safe cross 
sections for Sudakov-type processes~\cite{Sudakov,SoftGluonSummation}.  
For example, the splitting kernel 
$P_{ii}(x)$ describes the probability for a parton $i$ to split 
collinearly into a parton of the same species $i$, plus anything else, 
where the second parton $i$ retains a fraction $x$ of the 
longitudinal momentum of the first parton $i$.  
In the limit $x\to1$, this splitting kernel
is dominated by soft-gluon emission, and has the form,
\be
P_{ii}(x) \to { A(\alpha_s) \over (1-x)_+ } 
+ B(\alpha_s) \delta(1-x) + \ldots, 
\qquad  {\rm as}\ \ x\to 1,
\label{largexP}
\ee
where $A(\alpha_s)$ is related~\cite{KM}
to the soft (cusp) anomalous dimension
by,
\begin{equation}
 A(\alpha_s)  = {1\over 2} \, \gamma_K(\alpha_s) \,.
\end{equation}
The splitting kernel is related by a Mellin transform to the
anomalous dimensions of leading-twist operators of spin $j$,
\be
\gamma(j) \equiv - \int_0^1 dx \, x^{j-1} P(x).
\label{MellinDef}
\ee
Thus the soft anomalous dimension also controls the large-spin 
behavior of these anomalous dimensions~\cite{LeadingN}, 
\begin{equation}
\gamma(j) = {1\over2} \gamma_K(\alpha_s)\ ( \ln(j) + \gamma_e ) 
            - B(\alpha_s) + \Ord(\ln(j)/j)\,,
\label{gammajgammaK}
\end{equation}
where here we take $\gamma_e$ as Euler's constant.

KLOV~\cite{KLOV} have made a very interesting observation: the
anomalous dimensions of \SYM\ may be extracted directly from
the corresponding anomalous dimensions of QCD~\cite{MVV}, by keeping
terms of highest ``transcendentality''.  Recall that for the case
of the soft anomalous dimensions (large $j$ limit), the
transcendentality weight is simply $n$ for $\zeta_n$. (Although the
QCD anomalous dimensions are computed in the $\overline{\rm MS}$
regularization scheme, whereas for \SYM\ the DR~\cite{Siegel} or
FDH~\cite{FDH} schemes are needed to preserve supersymmetry, the
scheme-dependent terms drop out because they are of lower transcendentality.)
Although there is no proof of KLOV's prescription for extracting the
\SYM\ anomalous dimensions from QCD, there are good reasons to believe
that it is true~\cite{KLOV,TwoLoopSplitting}.  

Here we provide further evidence for the prescription, by 
confirming the large-spin behavior of the leading-twist 
anomalous dimension.  We compare the KLOV result, 
given in eqs.~(18)-(20) of ref.~\cite{KLOV},
against our evaluation of the same quantity from the
IR divergences of the three-loop four-point amplitude. 
Note that their normalization convention for anomalous
dimensions has an opposite overall sign from ours 
(which follows ref.~\cite{MVV}).
Also taking into account factors of 2 from the different 
$\alpha_s$ expansion parameter, and from \eqn{gammajgammaK},
we obtain from eqs.~(18)-(20) of ref.~\cite{KLOV},
\begin{eqnarray}
 \gamma_K^{(1)} &=& 4 \,\Nc \,, \nn\\
 \gamma_K^{(2)} &=& - 4 \zeta_2  \, \Nc^2 \,,  \\
 \gamma_K^{(3)} &=& 22 \,\zeta_4 \, \Nc^3 \,, \nn
\label{KLOVMVVgammaK}
\end{eqnarray}
which agrees perfectly with our results~(\ref{NcRescaledGammaG})
and (\ref{GammaValues}).

We remark that the {\it strong}-coupling, large-$\Nc$ limit of the
soft anomalous dimension $\gamma_K$ has been obtained, using
the AdS/CFT correspondence and classical 
supergravity methods~\cite{StrongCoupling}.  An approximate formula 
interpolating between the weak and strong-coupling limits has also been
constructed~\cite{KLV,KLOV}.

The coefficient ${\cal G}_0^{(l)}$, which controls the $1/\e$
singularity, may be extracted~\cite{EynckLaenenMagnea} from a 
fixed-order computation of the form factor at $l$ loops.  
For example, the two-loop quark form factor in QCD was computed 
in ref.~\cite{SudakovFormFactors}.
From eqs.~(21)--(22) of ref.~\cite{StermanTY}, if we follow the KLOV procedure
and keep the maximal transcendentality terms ($\zeta_3$ at two loops)
in order to convert the QCD results into \SYM\ results, we have,
\begin{eqnarray}
{\cal G}_0^\oneloop &=& 0\,, \nn \\
{\cal G}_0^\twoloop &=&  -\zeta_3\, \Nc^2 \,.
\end{eqnarray}
We have multiplied ${\cal G}_0^\twoloop$ in eq.~(22) of ref.~\cite{StermanTY}
by a factor of 4 to account for the different normalization conventions
used here.  These results agree with our 
\eqns{NcRescaledGammaG}{calGValues}.
Although the QCD form factors have not yet been computed at three loops,
we may use our results, together with the observation of KLOV,
to predict the leading-transcendentality contributions for QCD,
\begin{equation}
{\cal G}_0^{(3)} =  \biggl(4 \zeta_5 
                     + {10\over 3} \zeta_2 \zeta_3 \biggr) \Nc^3  \,,
\label{QCDpredict}
\end{equation}
after the group theory Casimirs have been set to the 
values $C_F = C_A = \Nc$. 
(At three loops, no other Casimirs can appear, so there are no 
subleading-color corrections to this leading-transcendentality 
prediction.)

\section{Conclusions and outlook}
\label{ConclusionSection}

In this paper we have provided strong evidence supporting the
conjecture~\cite{Iterate2} that the planar contributions to the
scattering amplitudes of \SYM\ possess an iterative structure.  This
result is in line with the growing body of evidence that gauge theory
amplitudes in general, and those of \SYM\ in particular, 
have a much simpler structure than had been anticipated.

Our evidence of iteration is based on a direct evaluation of the
planar three-loop four-point amplitude of \SYM.  The loop integrands for
this amplitude were obtained~\cite{BRY,BDDPR} using the unitarity 
method~\cite{NeqFourOneLoop,Fusing, UnitarityMachinery,OneLoopReview,%
TwoLoopSplitting}.  This method ensures that simple structures
uncovered at lower loop orders (including tree level)
in turn feed into higher loops.
(It also underlies much of the recent progress at one
loop~\cite{NewOneloop}.) 
In order to evaluate the required three-loop
integrals, we made use of important recent advances in multi-loop
integration~\cite{SmirnovDoubleBox,SmirnovVeretin,%
LoopIntegrationAdvance,SmirnovTripleBox}.
The integrals are expressed in terms of well-studied harmonic
polylogarithms~\cite{HPL,HPL2}, making it
straightforward to confirm the three-loop iteration.  
A rather intricate set of cancellations is required,
amongst the harmonic polylogarithms, and between different loop
integral types contributing to the amplitudes.

Using our explicitly computed four-point amplitudes as a springboard,
the known structure of infrared singularities to all loop
orders~\cite{MagneaSterman,StermanTY}, and the required factorization
properties of amplitudes, we constructed the ansatz for the
resummed $n$-point all-loop MHV amplitudes given in
\eqn{ExponentialResum}.  After subtracting the IR divergences, the
all-loop finite remainders (\ref{F0ResumFinal}) are given in terms of
known one-loop $n$-point finite remainders, as well as two
coefficients, one of which is the large-spin limit of the
leading-twist anomalous dimensions.

Very interestingly, the same set of leading-twist anomalous dimensions
has recently been linked to integrability of \SYM\ 
by Beisert, Kristjansen and Staudacher~\cite{MoreIntegrable,Staudacher}.
With the assumption of integrability, Staudacher~\cite{Staudacher} 
has reproduced the leading-twist anomalous dimensions at three loops
for spin $j$ up to 8.  These anomalous dimensions were previously
obtained by Kotikov, Lipatov, Onishchenko and Velizhanin~\cite{KLOV}
from the QCD results of Moch, Vermaseren and Vogt~\cite{MVV}.
(Quite recently, Staudacher's Bethe ansatz analysis has been extended to 
extremely high spins, the region relevant here, confirming the prediction
of KLOV for even values of $j$ up to 70~\cite{StaudacherPrivate}.)  
If one were able to
push this method to higher loop orders, and arbitrarily large spins,
it would give very directly the soft anomalous dimensions
appearing in our all-loop exponentiation of the MHV scattering
amplitudes.

Besides confirming the iterative structure of the scattering
amplitudes, our paper provides non-trivial confirmation of the form of
the three-loop divergences predicted by Sterman and
Tejeda-Yeomans~\cite{StermanTY}.  It also provides supporting evidence
for a number of ans\"atze appearing in a variety of papers.  In
particular, we confirm, in the high spin limit, the inspired ans\"atze
of KLOV, and (via KLOV) of Beisert, Kristjansen and Staudacher, 
for obtaining the leading-twist anomalous dimensions in \SYM.  
By making use of KLOV's link to QCD, via the
degree of transcendentality, our work also checks indirectly a small
piece of the three-loop splitting kernels in QCD, or equivalently the
anomalous dimensions of leading-twist operators, computed by Moch,
Vermaseren and Vogt~\cite{MVV}.  The integrand~\cite{BRY,BDDPR} used
in the computation of the planar three-loop four-point amplitude has
not been completely proven, but the match between its IR singularities
and the formulae of Sterman and Tejeda-Yeomans, plus the demonstration
of its iterative structure through the finite terms as $\e\to0$,
leaves little doubt as to its veracity.

The properties found here and in ref.~\cite{Iterate2} bring up the
possibility that the entire perturbative series of planar \SYM\ is
tractable.  The apparent simple structure of the MHV all-loop
amplitudes suggests that a loop-level twistor string interpretation
will be found~\cite{WittenTopologicalString,BerkovitsWitten}.  It
would be important to first identify the precise symmetry responsible
for this structure. A more complete understanding of the iterative
structure of the amplitudes should lead to important insights into
quantum field theory and the AdS/CFT correspondence.

\section*{Note added}

Since the first version of this paper came out, an interesting paper
has appeared~\cite{BC}, containing a technique for computing large classes
of terms for multi-loop \SYM\ amplitudes with many external legs,
which may shed further light on the iterative relations discussed here.
Also, the prediction~(\ref{QCDpredict}) for the 
leading-transcendentality terms in ${\cal G}_0^{(3)}$ for QCD
has now been confirmed~\cite{ThreeLoopQCDFormFactor}.

\section*{Acknowledgments}

We would like to thank Iosif Bena, Carola Berger, Eric D'Hoker, Gudrun
Heinrich, Per Kraus, Sven Moch, Emery Sokatchev, Matthias Staudacher,
George Sterman, Maria Elena Tejeda-Yeomans, and especially David
A. Kosower for helpful discussions.  We thank Pierpaolo Mastrolia for
providing us with his tables of harmonic polylogarithms.  We also
thank Academic Technology Services at UCLA for computer support.  This
research was supported by the US Department of Energy under contracts
DE--FG03--91ER40662 and DE--AC02--76SF00515. The work of V.S. was
supported by the Russian Foundation for Basic Research through project
05-02-17645 and by DFG Mercator Grant No.~Ha 202/110-1.

\appendix

\section{Harmonic polylogarithms}
\label{HarmonicPolyLogAppendix}

We express the amplitudes in terms of harmonic
polylogarithms~\cite{HPL}, which are generalizations of ordinary
polylogarithms~\cite{Lewin}.  Here we briefly summarize some salient
properties.  A more complete discussion is given in ref.~\cite{HPL}.
Recipes for numerically evaluating harmonic polylogarithms may be found in 
ref.~\cite{HPL2}.

The weight $n$ harmonic polylogarithms 
$H_{a_1 a_2 \ldots a_n}(x) \equiv H(a_1,a_2,\ldots,a_n;x)$, 
with $a_i \in \{1,0,-1\}$, are defined recursively by,
\begin{equation}
H_{a_1 a_2 \ldots a_n}(x) 
= \int_0^x \dd t \,  f_{a_1}(t) H_{a_2 \ldots a_n}(t)\,,
\end{equation}
where
\begin{eqnarray}
f_{\pm 1}(x) &=& {1 \over 1 \mp x} \,, \qquad f_{0}(x) = {1\over x} \,, \\
H_{\pm 1}(x) &=& \mp \ln(1\mp x) \,, \qquad H_{0}(x)= \ln x \,,
\end{eqnarray}
and at least one of the indices $a_i$ is non-zero.
For all $a_i=0$, one has
\be
H_{0,0,\ldots,0}(x) = \frac{1}{n!}\ln^n x\;.
\ee

If a given harmonic polylogarithm involves only parameters $a_i=0$ and
$1$, and the number of these parameters (the weight) is less than or equal 
to four, it can be expressed \cite{HPL} in terms of the standard
polylogarithms~\cite{Lewin},
\begin{eqnarray}
\Li_n(z) &=& \sum_{j=1}^\infty {z^j \over j^n} =
\int_0^z {\dd t\over t} \Li_{n-1}(t) \,, \nonumber\\
\Li_2(z) &=& -\int_0^z {\dd t\over t} \ln(1-t) \,,
\label{PolyLogDef}
\end{eqnarray}
with $n=2,3,4$, and where $z$ may take the values $x$, $1/(1-x)$, 
or $-x/(1-x)$.  (For $n<4$, not all of these values are required,
due to identities.)
Here we need only $a_i \in \{0,1\}$, but weights up to six.
In the Euclidean region for the planar four-point process,
namely $s<0$, $t<0$, $u>0$, with the identification $x=-t/s$,
the argument $x$ of the harmonic polylogarithms will be negative.

The harmonic polylogarithms are not all independent; 
they are related by sets of identities~\cite{HPL}.  
One set of identities, derived using integration by parts,
\be
H_{a_1 a_2 \ldots a_p 0}(x) 
= \ln x \, H_{a_1 a_2 \ldots a_p}(x)
         - H_{0 a_1 a_2 \ldots a_p}(x)
         - H_{a_1 0 a_2 \ldots a_p}(x)
         - \cdots - H_{a_1 a_2 \ldots 0 a_p}(x) \,,
\label{RemoveTrailingZeroes}
\ee
allows one to remove trailing zeroes from the string of parameters 
$a_i$.  The remaining $H_{a_1 a_2 \ldots a_n}(x)$ with $a_n=1$
are well-behaved as $x \to 0$; in fact they all vanish there.

Because the integrals appear in the \SYM\ amplitudes with 
arguments $(s,t)$ and $(t,s)$, we need a set of identities relating
harmonic polylogarithms with argument $x = -t/s$ to those with argument
$y = -s/t = 1/x$.  As explained in ref.~\cite{HPL} (see the discussion near
eqs.~(55) of that reference), we may construct the required set of
identities by induction on the weight of the harmonic polylogarithms.
For the first few weights, in the region $-1 \ge x \ge 0$, 
and letting $L=\ln(s/t) = \ln(-1/x)$,
we have, for example,
\begin{eqnarray}
H_{1}(y) &=& H_{1}(x) - L \,, \nonumber \\
H_{0,1}(y) &=&  - H_{0,1}(x) - {1\over 2} L^2 - {\pi^2 \over 6}\,, 
    \nonumber\\
H_{1,1}(y) &=&  H_{1,1}(x) - H_{1}(x) L + {1\over 2} L^2 \,, 
    \nonumber\\
H_{0,0,1}(y) &=&  H_{0,0,1} (x) - {\pi^2\over 6} \, L - {1\over 6} L^3 \,,
              \nn \\
H_{0,1,1}(y) &=&  H_{0, 0, 1}(x) - H_{0, 1, 1}(x) + H_{0, 1}(x) L  
                  + {1\over 6} L^3 +  \zeta_3 \,, \nn\\
H_{1,0,1}(y) &=&  - 2 H_{0,0,1}(x) + 2 H_{0,1,1}(x)  - 2 H_{0,1}(x) L  
 - {\pi^2\over 6} H_{1}(x) - H_{1}(x) H_{0, 1}(x)  \nn\\
&& \null      -  {1\over 2} H_{1}(x) L^2 - {1\over 3} L^3 + {\pi^2 \over 6} L  
     + H_{0, 1}(x) L + {1\over2} L^3  - 2\, \zeta_3 \,, \nn\\
H_{1,1,1}(y) &=&  H_{1,1,1}(x) - H_{1,1}(x) L  
 + {1\over2} H_{1}(x) L^2 - {1\over 6} L^3 \,.
\label{invertx}
\end{eqnarray}
%


\section{Integrals appearing in four-point amplitudes}
\label{IntegralsAppendix}

In this appendix we collect various integrals that are needed
as well as their values in terms of harmonic polylogarithms.
We quote the results in the Euclidean ($u$-channel) region, $s,t <0$.
The analytic continuation to other physical regions is discussed
in refs.~\cite{HPL,HPL2}. 

\subsection{One-loop integrals}
\label{OneLoopBoxIntegralSubsection}

%
\begin{figure}[t]
\centerline{\epsfxsize 1.5 truein \epsfbox{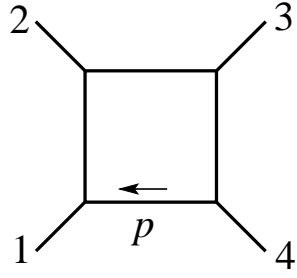}}
\caption{The one-loop box integral.}
\label{SingleBoxFigure}
\end{figure}

Consider the (conveniently normalized) one-loop scalar box integral,
depicted in \fig{SingleBoxFigure},
\begin{equation}
 I_4^\oneloop(s,t) =
-i e^{\e \gamma} \pi^{-d/2}
\int   {\dd^d p \over p^2 \, (p-k_1)^2 \, (p-k_1 - k_2)^2 (p+k_4)^2} \,.
\label{OneLoopBoxDef}
\end{equation}
The value of this integral, with $x = -t/s, L = \ln(s/t)$, is
\begin{equation}
I_4^\oneloop(s,t) =  - {1 \over (-s)^{1+\e} t}
                \sum_{j=-4}^{2} {c_j(x, L) \over \e^j} \,,
\label{OneLoopBoxResult}
\end{equation}
\def\hs{\hspace*{1cm}}
with
\begin{eqnarray}
c_2 & = &
 4
\,, \nn \\
c_1 &=&
  2 L
\,, \nn \\
c_0 & = &
-\frac{4}{3}\, \pi^2
\,, \nn \\
c_{-1} & =&
 \pi^2 \H{1}{(x)} + 2 \H{0, 0, 1}{(x)} - \frac{7}{6} \pi^2  L
 + 2 \H{0, 1}{(x)} L + \H{1}{(x)} L^2 - \frac{1}{3} L^3
 -  \frac{34}{3} \zeta_3
\,, \nn \\
c_{-2} & =&
  - 2 \H{1, 0, 0, 1}{(x)}
  - 2 \H{0, 0, 1, 1}{(x)}
  - 2 \H{0, 1, 0, 1}{(x)}
  - 2 \H{0, 0, 0, 1}{(x)}
  - 2 \H{0, 1, 1}{(x)} L
\nn \\
&& \null
  - 2 \H{1, 0, 1}{(x)} L
  + \H{0, 1}{(x)} L^2
  - \H{1, 1}{(x)} L^2
  + {2\over 3} \H{1}{(x)} L^3
  - {1\over 6} L^4
\nn \\
&& \null
  - \pi^2 \H{1, 1}{(x)}
  + \pi^2 \H{1}{(x)} L
  - {1\over 2} \pi^2 L^2
  + 2 \H{1}{(x)} \zeta_3
  - {20\over 3} L \zeta_3  - {41\over360}  \pi^4
\,, \nn \\
c_{-3} & =&
      2 \H{1, 0, 0, 0, 1}{(x)}
    + 2 \H{1, 0, 0, 1, 1}{(x)}
    + 2 \H{1, 0, 1, 0, 1}{(x)}
    + 2 \H{1, 1, 0, 0, 1}{(x)}
    + 2 \H{0, 0, 0, 0, 1}{(x)}
\nn \\
&& \null
    + 2 \H{0, 0, 0, 1, 1}{(x)}
    + 2 \H{0, 0, 1, 0, 1}{(x)}
    + 2 \H{0, 0, 1, 1, 1}{(x)}
    + 2\H{0, 1, 0, 0, 1}{(x)}
    + 2 \H{0, 1, 0, 1, 1}{(x)}
\nn \\
&& \null
    + 2 \H{0, 1, 1, 0, 1}{(x)}
    + 2 \H{0, 1, 1, 1}{(x)} L
    + 2 \H{1, 0, 1, 1}{(x)} L
    + 2 \H{1, 1, 0, 1}{(x)} L
    -   \H{0, 1, 1}{(x)} L^2
\nn \\
&& \null
    -   \H{1, 0, 1}{(x)} L^2
    +   \H{1, 1, 1}{(x)} L^2
    + {1\over 3} \H{0, 1}{(x)} L^3
    - {2\over3} \H{1, 1}{(x)} L^3
    + {1\over4} \H{1}{(x)} L^4
    - {1\over 20} L^5
\nn \\
&& \null
    - {1\over 6} \pi^2 \H{0, 0, 1}{(x)}
    + \pi^2 \H{1, 1, 1}{(x)}
    - {1\over 6} \pi^2 \H{0, 1}{(x)} L
    -  \pi^2 \H{1, 1}{(x)} L
    + {5 \over 12} \pi^2 \H{1}{(x)} L^2
\nn \\
&& \null
    - {5\over36} \pi^2 L^3
    + {59\over18} \pi^2 \zeta_3
    - 2 \H{1, 1}{(x)} \zeta_3
    + 2 \H{1}{(x)} L \zeta_3
    -  L^2 \zeta_3
\nn \\
&& \null
    - {7 \over 144} \pi^4 L
     - {1\over 60} \pi^4 \H{1}{(x)}
    - {134\over5} \zeta_5
\,, \nn \\
c_{-4} & = &
   - 2 \H{0, 0, 0, 0, 0, 1}{(x)}
   - 2 \H{0, 0, 0, 0, 1, 1}{(x)}
   - 2 \H{0, 0, 0, 1, 0, 1}{(x)}
   - 2 \H{0, 0, 0, 1, 1, 1}{(x)}
   - 2 \H{0, 0, 1, 0, 0, 1}{(x)}
\nn \\
&& \null
   - 2 \H{0, 0, 1, 0, 1, 1}{(x)}
   - 2 \H{0, 0, 1, 1, 0, 1}{(x)}
   - 2 \H{0, 0, 1, 1, 1, 1}{(x)}
   - 2 \H{0, 1, 0, 0, 0, 1}{(x)}
   - 2 \H{0, 1, 0, 0, 1, 1}{(x)}
\nn \\
&& \null
   - 2 \H{0, 1, 0, 1, 0, 1}{(x)}
   - 2 \H{0, 1, 0, 1, 1, 1}{(x)}
   - 2 \H{0, 1, 1, 0, 0, 1}{(x)}
   - 2 \H{0, 1, 1, 0, 1, 1}{(x)}
   - 2 \H{0, 1, 1, 1, 0, 1}{(x)}
\nn \\
&& \null
   - 2 \H{1, 0, 0, 0, 0, 1}{(x)}
   - 2 \H{1, 0, 0, 0, 1, 1}{(x)}
   - 2 \H{1, 0, 0, 1, 0, 1}{(x)}
   - 2 \H{1, 0, 0, 1, 1, 1}{(x)}
   - 2 \H{1, 0, 1, 0, 0, 1}{(x)}
\nn \\
&& \null
   - 2 \H{1, 0, 1, 0, 1, 1}{(x)}
   - 2 \H{1, 0, 1, 1, 0, 1}{(x)}
   - 2 \H{1, 1, 0, 0, 0, 1}{(x)}
   - 2 \H{1, 1, 0, 0, 1, 1}{(x)}
   - 2 \H{1, 1, 0, 1, 0, 1}{(x)}
\nn \\
&& \null
   - 2 \H{1, 1, 1, 0, 0, 1}{(x)}
   - 2 \H{0, 1, 1, 1, 1}{(x)} L
   - 2 \H{1, 0, 1, 1, 1}{(x)} L
   - 2 \H{1, 1, 0, 1, 1}{(x)} L
   - 2 \H{1, 1, 1, 0, 1}{(x)} L
\nn \\
&& \null
   + \H{0, 1, 1, 1}{(x)} L^2
   + \H{1, 0, 1, 1}{(x)} L^2
   + \H{1, 1, 0, 1}{(x)} L^2
   - \H{1, 1, 1, 1}{(x)} L^2
   - \frac{1}{60} \pi^4 \H{1}{(x)} L
\nn \\
&& \null
   + \frac{1}{6} \pi^2 \H{0, 1, 1}{(x)} L
   + \frac{1}{6} \pi^2 \H{1, 0, 1}{(x)} L
   +  \pi^2 \H{1, 1, 1}{(x)} L
   + \frac{1}{120}\pi^4 L^2
   - \frac{1}{12} \pi^2 \H{0, 1}{(x)} L^2
\nn \\
&& \null
   - {5 \over 12} \pi^2 \H{1, 1}{(x)} L^2
   + \frac{1}{9} \pi^2 \H{1}{(x)} L^3
   - \frac{1}{3} \H{0, 1, 1}{(x)} L^3
   - \frac{1}{3} \H{1, 0, 1}{(x)} L^3
   + {2\over 3} \H{1, 1, 1}{(x)} L^3
\nn \\
&& \null
   - {1\over 36} \pi^2 L^4
   + {1\over 12} \H{0, 1}{(x)} L^4
   - {1\over 4} \H{1, 1}{(x)} L^4
   + {1\over 15} \H{1}{(x)} L^5
   - \frac{1}{90} L^6
   + \frac{1}{60}\pi^4 \H{1, 1}{(x)}
\nn \\
&& \null
   + \frac{1}{6} \pi^2 \H{0, 0, 0, 1}{(x)}
   + \frac{1}{6} \pi^2 \H{0, 0, 1, 1}{(x)}
   + \frac{1}{6} \pi^2 \H{0, 1, 0, 1}{(x)}
   + \frac{1}{6} \pi^2 \H{1, 0, 0, 1}{(x)}
   - \pi^2 \H{1, 1, 1, 1}{(x)}
\nn \\
&& \null
   - \frac{5}{2} \pi^2 \H{1}{(x)} \zeta_3
   - \frac{14}{3} \H{0, 0, 1}{(x)} \zeta_3
   + 2 \H{1, 1, 1}{(x)} \zeta_3
   + \frac{26}{9} \pi^2 L \zeta_3
   - \frac{14}{3}  \H{0, 1}{(x)} L \zeta_3
\nn \\
&& \null
   - 2 \H{1, 1}{(x)} L \zeta_3
   - \frac{4}{3} \H{1}{(x)} L^2 \zeta_3
   + \frac{4}{9} L^3 \zeta_3
   + \frac{140}{9} \zeta_3^2
   + 2 \H{1}{(x)} \zeta_5
   - \frac{72}{5} L \zeta_5
   + \frac{1}{2160} \pi^6
\! . \label{ExplicitOneloopBox}
\end{eqnarray}
%

\subsection{Two-loop integrals}
\label{TwoLoopBoxIntegralSubsection}

%
\begin{figure}[t]
\centerline{\epsfxsize 2. truein \epsfbox{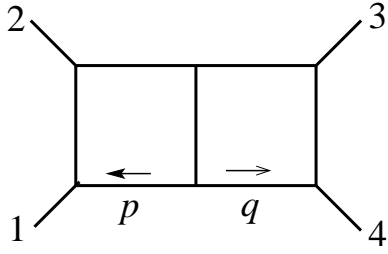}}
\caption{The two-loop double-box integral.}
\label{DoubleBoxFigure}
\end{figure}

The two-loop planar scalar double-box integral depicted 
in~\fig{DoubleBoxFigure} is
\begin{equation}
I_4^{\twoloop}(s,t) =
(-i e^{\e \gamma} \pi^{-d/2})^2
\int
 {\dd^d p \, \dd^d q \,\over p^2 \, (p - k_1)^2 \,(p - k_1 - k_2)^2 \,
 (p + q)^2 q^2 \,
        (q-k_4)^2 \, (q - k_3 - k_4)^2 }\,.
\label{TwoLoopBoxDef}
\end{equation}
This integral was first evaluated in ref.~\cite{SmirnovDoubleBox} through
$\Ord(\e^0)$, as required in NNLO calculations.  Here we need the
integral through $\Ord(\e^2)$.  The calculation performed
in ref.~\cite{SmirnovDoubleBox} was not optimal because the
starting point was a fivefold MB representation. On the other
hand, it is possible to derive an appropriate fourfold representation,
as was demonstrated in ref.~\cite{SDBox4MB}
(see also Chapter~4 of ref.~\cite{Buch}).
The corresponding evaluation can be generalized straightforwardly
to obtain the next two orders of the expansion in $\ep$. Let us stress
that this evaluation is much simpler than the evaluation of the
triple boxes up to $\ep^0$.

Our result through $\Ord(\ep^2)$ is,
\begin{equation}
I_4^{\twoloop}(s,t)=
- \frac{1}{(-s)^{2+2\ep} t}
\;
 \sum_{j=-2}^4 \frac{ c_j(x,L)}{\ep^j}\;,
\label{TwoLoopPDBResult}
\end{equation}
where 
$x=-t/s$, $L=\ln(s/t)$, and
\def\hs{\hspace*{-81mm}}
\begin{eqnarray}
\;\;\;c_4=
-4
\,, \;\;\;
%
%
c_3=
- 5 L
\,, \;\;\;
%
%
c_2=
 -2 L^2 + {5\over 2} \pi^2
\,, \;\;\;
\nn \\ &&  \hspace*{-81mm}
c_1=
4 \left[-L \H{0, 1}{(x)} - \H{0, 0, 1}{(x)}\right] -
2 \left(L^2 + \pi^2\right) \H{1}{(x)} + {2\over 3} L^3
+ {11\over 2} L \pi^2 + {65\over 3} \zeta_3
\,,
\nn \\ &&  \hspace*{-81mm}
c_0=
4 \left[11 \H{0, 0, 0, 1}{(x)} + \H{0, 0, 1, 1}{(x)} + \H{0, 1, 0, 1}{(x)} +
\H{1, 0, 0, 1}{(x)}\right]
\nn \\ &&  \hs
+ 4 L {} \left[6 \H{0, 0, 1}{(x)} + \H{0, 1, 1}{(x)}
                 + \H{1, 0, 1}{(x)}\right] +
2 L^2\left[\H{0, 1}{(x)} + \H{1, 1}{(x)} \right]
\nn \\ &&  \hs
+ {2\over 3} \pi^2 \left[10 \H{0, 1}{(x)} + 3 \H{1, 1}{(x)}\right]
+ {2\over 3} \H{1}{(x)} \left[-4 L^3 - 5 L  \pi^2 - 6 \zeta_3\right]
\nn \\ &&  \hs
+ {4\over 3} L^4 +
6 \pi^2 L^2 + {29\over 30} \pi^4 + {88\over 3} \zeta_3 L
\,,
\nn \\ &&  \hspace*{-81mm}
c_{-1}=
-4 \left[28 \H{0, 0, 0, 0, 1}{(x)} + 29 \H{0, 0, 0, 1, 1}{(x)} +
24 \H{0, 0, 1, 0, 1}{(x)} + \H{0, 0, 1, 1, 1}{(x)}  \right.
\nn \\ &&  \hs
+ 19 \H{0, 1, 0, 0, 1}{(x)}+ \H{0, 1, 0, 1, 1}{(x)} +
\H{0, 1, 1, 0, 1}{(x)} + 14 \H{1, 0, 0, 0, 1}{(x)}
\nn \\ &&  \hs \left.
+ \H{1, 0, 0, 1, 1}{(x)} + \H{1, 0, 1, 0, 1}{(x)} +
\H{1, 1, 0, 0, 1}{(x)}\right]
\nn \\ &&  \hs
- 4 L {}\left[18 \H{0, 0, 1, 1}{(x)} + 13 \H{0, 1, 0, 1}{(x)}
     + \H{0, 1, 1, 1}{(x)} +
8 \H{1, 0, 0, 1}{(x)} + \H{1, 0, 1, 1}{(x)} + \H{1, 1, 0, 1}{(x)}\right]
\nn \\ &&  \hs
+ 2 L^2 \left[12 \H{0, 0, 1}{(x)} - 7 \H{0, 1, 1}{(x)} - 2 \H{1, 0, 1}{(x)} -
\H{1, 1, 1}{(x)}\right]
\nn \\ &&  \hs
+ {2\over 3} \pi^2 \left[\H{0, 0, 1}{(x)} -
28 \H{0, 1, 1}{(x)} - 13 \H{1, 0, 1}{(x)} - 3 \H{1, 1, 1}{(x)}\right]
\nn \\ &&  \hs
+ {8\over 3} L^3 \left[2 \H{0, 1}{(x)} + \H{1, 1}{(x)}\right]
+ {2\over 3} L \pi^2 \left[18 \H{0, 1}{(x)} + 5 \H{1, 1}{(x)}\right]
\nn \\ &&  \hs
+ 72 \zeta_3\H{0, 1}{(x)} - {1\over 18} \left[36 L^4 + 78 L^2 \pi^2
+ 17 \pi^4\right] \H{1}{(x)} -
4 \zeta_3\left[-7 L \H{1}{(x)} - \H{1, 1}{(x)}\right]
\nn \\ &&  \hs
+ {14\over 15} L^5 + {13\over 3} \pi^2 L^3
+ {46\over 3}\zeta_3 L^2 + {211\over 120}  \pi^4 L
- {73\over 6} \pi^2 \zeta_3 + {383\over 5} \zeta_5
\,,
\nn \\ &&  \hspace*{-81mm}
c_{-2}=
4 \left[68 \H{0, 0, 0, 0, 0, 1}{(x)} + 76 \H{0, 0, 0, 0, 1, 1}{(x)} +
66 \H{0, 0, 0, 1, 0, 1}{(x)} + 65 \H{0, 0, 0, 1, 1, 1}{(x)}
\right.
\nn \\ &&  \hs
+ 56 \H{0, 0, 1, 0, 0, 1}{(x)} + 60 \H{0, 0, 1, 0, 1, 1}{(x)} +
48 \H{0, 0, 1, 1, 0, 1}{(x)} + \H{0, 0, 1, 1, 1, 1}{(x)}
\nn \\ &&  \hs
+ 46 \H{0, 1, 0, 0, 0, 1}{(x)} + 55 \H{0, 1, 0, 0, 1, 1}{(x)} +
43 \H{0, 1, 0, 1, 0, 1}{(x)} + \H{0, 1, 0, 1, 1, 1}{(x)}
\nn \\ &&  \hs
+ 31 \H{0, 1, 1, 0, 0, 1}{(x)} + \H{0, 1, 1, 0, 1, 1}{(x)} +
\H{0, 1, 1, 1, 0, 1}{(x)} + 36 \H{1, 0, 0, 0, 0, 1}{(x)}
\nn \\ &&  \hs
+ 50 \H{1, 0, 0, 0, 1, 1}{(x)} + 38 \H{1, 0, 0, 1, 0, 1}{(x)} +
\H{1, 0, 0, 1, 1, 1}{(x)} + 26 \H{1, 0, 1, 0, 0, 1}{(x)}
\nn \\ &&  \hs
+ \H{1, 0, 1, 0, 1, 1}{(x)} + \H{1, 0, 1, 1, 0, 1}{(x)} +
14 \H{1, 1, 0, 0, 0, 1}{(x)} + \H{1, 1, 0, 0, 1, 1}{(x)}
\nn \\ &&  \hs \left.
+ \H{1, 1, 0, 1, 0, 1}{(x)} + \H{1, 1, 1, 0, 0, 1}{(x)}\right] +
4 L {} \left[42 \H{0, 0, 1, 1, 1}{(x)} + 37 \H{0, 1, 0, 1, 1}{(x)}
\right.
\nn \\ &&  \hs
+ 25 \H{0, 1, 1, 0, 1}{(x)} + \H{0, 1, 1, 1, 1}{(x)} +
32 \H{1, 0, 0, 1, 1}{(x)} + 20 \H{1, 0, 1, 0, 1}{(x)} +
\H{1, 0, 1, 1, 1}{(x)}
\nn \\ &&  \hs \left.
+ 8 \H{1, 1, 0, 0, 1}{(x)} +
\H{1, 1, 0, 1, 1}{(x)} + \H{1, 1, 1, 0, 1}{(x)}\right] -
2 L^2 \left[36 \H{0, 0, 1, 1}{(x)} + 26 \H{0, 1, 0, 1}{(x)}
\right.
\nn \\ &&  \hs
\left.
- 19 \H{0, 1, 1, 1}{(x)} + 16 \H{1, 0, 0, 1}{(x)} -
14 \H{1, 0, 1, 1}{(x)} - 2 \H{1, 1, 0, 1}{(x)} -
\H{1, 1, 1, 1}{(x)}\right]
\nn \\ &&  \hs
- {2\over 3} \pi^2 \left[17 \H{0, 0, 0, 1}{(x)} +
\H{0, 0, 1, 1}{(x)} + 6 \H{0, 1, 0, 1}{(x)} - 64 \H{0, 1, 1, 1}{(x)}
\right.
\nn \\ &&  \hs
\left.
+ 11 \H{1, 0, 0, 1}{(x)} - 49 \H{1, 0, 1, 1}{(x)}
- 13 \H{1, 1, 0, 1}{(x)} - 3 \H{1, 1, 1, 1}{(x)}\right]
\nn \\ &&  \hs
+ {8\over 3} L^3 \left[6 \H{0, 0, 1}{(x)} - 8 \H{0, 1, 1}{(x)}
- 3 \H{1, 0, 1}{(x)} - \H{1, 1, 1}{(x)}\right]
\nn \\ &&  \hs
- {2\over 3}\pi^2 L {} \left[6 \H{0, 0, 1}{(x)} +
54 \H{0, 1, 1}{(x)} + 29 \H{1, 0, 1}{(x)} + 5 \H{1, 1, 1}{(x)}\right]
\nn \\ &&  \hs
- {4\over 3} \zeta_3 \left[40 \H{0, 0, 1}{(x)} + 90 \H{0, 1, 1}{(x)} +
75 \H{1, 0, 1}{(x)} + 3 \H{1, 1, 1}{(x)}\right]
+ {2\over 3} L^4 \left[7 \H{0, 1}{(x)} + 3 \H{1, 1}{(x)}\right]
\nn \\ &&  \hs
+ {1\over 3} \pi^2 L^2  \left[33 \H{0, 1}{(x)} + 13 \H{1, 1}{(x)}\right]
+ {1\over 90} \pi^4 \left[129 \H{0, 1}{(x)} + 85 \H{1, 1}{(x)}\right]
\nn \\ &&  \hs
+ {4\over 3} \zeta_3 L {}\left[59 \H{0, 1}{(x)} - 21 \H{1, 1}{(x)}\right]
\nn \\ &&  \hs
+ {1\over 45}  \left[- 48 L^5 -
160 L^3 \pi^2 - 55 L \pi^4 + 1140 L^2 \zeta_3 + 240 \pi^2 \zeta_3 -
720 \zeta_5\right] \H{1}{(x)}
\nn \\ &&  \hs
+ {4\over 9} L^6 + {7\over 3} \pi^2 L^4
+ {8\over 9} \zeta_3 L^3 + {19\over 12}\pi^4 L^2
- {98\over 3} \zeta_3 \pi^2 L
\nn \\ &&  \hs
+ 80 \zeta_5 L + {2357\over 15120} \pi^6 - {275\over 9} \zeta_3^2
\,. \label{TwoLoopBoxValue}
\end{eqnarray}
Through $\Ord(\e^0)$ this corresponds to the results of
ref.~\cite{SmirnovDoubleBox}.

It is also possible to derive differential equations obeyed by
the planar two-loop box integral~\cite{SmirnovVeretin}.
The differential equations couple $I_4^{\twoloop}(s,t)$ to 
a second master two-loop box integral.  In ref.~\cite{SmirnovVeretin}
these results were used to obtain the second integral, and to
check $I_4^{\twoloop}(s,t)$ through order $\e^0$.  
We have used the same differential equations to check the 
result~(\ref{TwoLoopBoxValue}) through the required order, $\e^2$,
up to a constant.  The order $\e^2$ constant was checked numerically.

\subsection{Three-loop integrals}
\label{ThreeLoopBoxIntegralSubsection}

The three-loop ladder integral depicted in \fig{TripleBoxFigure}(a)
and defined in \eqn{ThreeLoopIntegralA}
has been evaluated in ref.~\cite{SmirnovTripleBox}, with the result,
\begin{equation}
I_4^{\threeloop a} (s,t)=
-\frac{1}{s^3 (-t)^{1+3\ep}}
\;
 \sum_{j=0}^6 \frac{ c_j(x,L)}{\ep^j}\;,
\label{ThreeLoopaResult}
\end{equation}
where 
$x=-t/s$, $L=\ln(s/t)$, and
\def\hs{\hspace*{-74mm}}
\begin{eqnarray}
c_6=
\frac{16}{9}
\,, \;\;\;
%
%
c_5=
-\frac{5}{3} L
\,, \;\;\;
%
%
c_4=
-\frac{3}{2}\pi^2
\,, \;\;\;
\nn \\ &&  \hspace*{-74mm}
c_3=
3 (\H{0, 0, 1}{(x)} + L \H{0, 1}{(x)}) + \frac{3}{2} (L^2 + \pi^2) \H{1}{(x)} -
\frac{11}{12} \pi^2  L - \frac{131}{9} \zeta_3
\,,
\nn \\ &&  \hspace*{-74mm}
c_2=
-3 \left(17 \H{0, 0, 0, 1}{(x)} + \H{0, 0, 1, 1}{(x)} + \H{0, 1, 0, 1}{(x)} +
          \H{1, 0, 0, 1}{(x)}\right)
\nn \\ &&  \hs
- L {} \left(37 \H{0, 0, 1}{(x)}
          + 3  \H{0, 1, 1}{(x)} +  3  \H{1, 0, 1}{(x)}\right)
    - \frac{3}{2} (L^2 + \pi^2) \H{1, 1}{(x)}
\nn \\ &&  \hs
    - \left( \frac{23}{2} L^2 + 8 \pi^2 \right) \H{0, 1}{(x)}
    - \left(\frac{3}{2} L^3 + \pi^2 L - 3 \zeta_3 \right)\H{1}{(x)}
    + \frac{49}{3} \zeta_3 L - \frac{1411}{1080} \pi^4
\,,
\nn \\ &&  \hspace*{-74mm}
c_1=
3 \left( 81 \H{0, 0, 0, 0, 1}{(x)} + 41 \H{0, 0, 0, 1, 1}{(x)}
+ 37 \H{0, 0, 1, 0, 1}{(x)}
+ \H{0, 0, 1, 1, 1}{(x)} \right.
\nn \\ &&  \hs
+ 33 \H{0, 1, 0, 0, 1}{(x)} + \H{0, 1, 0, 1, 1}{(x)} + \H{0, 1, 1, 0, 1}{(x)}
+ 29 \H{1, 0, 0, 0, 1}{(x)}
\nn \\ &&  \hs \left.
+ \H{1, 0, 0, 1, 1}{(x)} + \H{1, 0, 1, 0, 1}{(x)} + \H{1, 1, 0, 0, 1}{(x)}\right)
+ L {}\left(177 \H{0, 0, 0, 1}{(x)} + 85 \H{0, 0, 1, 1}{(x)} \right.
\nn \\ &&  \hs \left.
+ 73 \H{0, 1, 0, 1}{(x)}
+ 3 \H{0, 1, 1, 1}{(x)} + 61 \H{1, 0, 0, 1}{(x)} + 3 \H{1, 0, 1, 1}{(x)}
+ 3 \H{1, 1, 0, 1}{(x)}\right)
\nn \\ &&  \hs
+ \left(\frac{119}{2} L^2 + \frac{139}{12} \pi^2\right) \H{0, 0, 1}{(x)}
+ \left(\frac{47}{2} L^2 + 20 \pi^2\right) \H{0, 1, 1}{(x)}
\nn \\ &&  \hs
+ \left(\frac{35}{2} L^2 + 14 \pi^2\right) \H{1, 0, 1}{(x)}
+ \frac{3}{2}\left(L^2 + \pi^2\right) \H{1, 1, 1}{(x)}
\nn \\ &&  \hs
+ \left(\frac{23}{2} L^3 + \frac{83}{12}\pi^2 L
- 96 \zeta_3 \right)\H{0, 1}{(x)}
+ \left(\frac{3}{2} L^3 + \pi^2 L - 3 \zeta_3 \right) \H{1, 1}{(x)}
\nn \\ &&  \hs
+ \left(\frac{9}{8} L^4 + \frac{25}{8} \pi^2 L^2 - 58 \zeta_3  L
+ \frac{13}{8} \pi^4\right) \H{1}{(x)}
- \frac{503}{1440}\pi^4 L
+ \frac{73}{4} \pi^2 \zeta_3  - \frac{301}{15} \zeta_5
\,,
\nn \\ &&  \hspace*{-74mm}
c_0=
-\left( 951 \H{0, 0, 0, 0, 0, 1}{(x)} + 819 \H{0, 0, 0, 0, 1, 1}{(x)}
+ 699 \H{0, 0, 0, 1, 0, 1}{(x)} + 195 \H{0, 0, 0, 1, 1, 1}{(x)}
\right.
\nn \\ &&  \hs
+ 547 \H{0, 0, 1, 0, 0, 1}{(x)} + 231 \H{0, 0, 1, 0, 1, 1}{(x)}
+ 159 \H{0, 0, 1, 1, 0, 1}{(x)} + 3 \H{0, 0, 1, 1, 1, 1}{(x)}
\nn \\ &&  \hs
+ 363 \H{0, 1, 0, 0, 0, 1}{(x)} + 267 \H{0, 1, 0, 0, 1, 1}{(x)}
+ 195 \H{0, 1, 0, 1, 0, 1}{(x)} + 3 \H{0, 1, 0, 1, 1, 1}{(x)}
\nn \\ &&  \hs
+ 123 \H{0, 1, 1, 0, 0, 1}{(x)} + 3 \H{0, 1, 1, 0, 1, 1}{(x)}
+ 3 \H{0, 1, 1, 1, 0, 1}{(x)} + 147 \H{1, 0, 0, 0, 0, 1}{(x)}
\nn \\ &&  \hs
+ 303 \H{1, 0, 0, 0, 1, 1}{(x)} + 231 \H{1, 0, 0, 1, 0, 1}{(x)}
+ 3 \H{1, 0, 0, 1, 1, 1}{(x)} + 159 \H{1, 0, 1, 0, 0, 1}{(x)}
\nn \\ &&  \hs
+ 3 \H{1, 0, 1, 0, 1, 1}{(x)} + 3 \H{1, 0, 1, 1, 0, 1}{(x)}
+ 87 \H{1, 1, 0, 0, 0, 1}{(x)} + 3 \H{1, 1, 0, 0, 1, 1}{(x)}
\nn \\ &&  \hs \left.
+ 3 \H{1, 1, 0, 1, 0, 1}{(x)} + 3 \H{1, 1, 1, 0, 0, 1}{(x)}
\right)
\nn \\ &&  \hs
- L {}\left(729 \H{0, 0, 0, 0, 1}{(x)} + 537 \H{0, 0, 0, 1, 1}{(x)}
  + 445 \H{0, 0, 1, 0, 1}{(x)} + 133 \H{0, 0, 1, 1, 1}{(x)}\right.
\nn \\ &&  \hs
  + 321 \H{0, 1, 0, 0, 1}{(x)} + 169 \H{0, 1, 0, 1, 1}{(x)}
  + 97 \H{0, 1, 1, 0, 1}{(x)} + 3 \H{0, 1, 1, 1, 1}{(x)}
\nn \\ &&  \hs
  + 165 \H{1, 0, 0, 0, 1}{(x)} + 205 \H{1, 0, 0, 1, 1}{(x)}
  + 133 \H{1, 0, 1, 0, 1}{(x)} + 3 \H{1, 0, 1, 1, 1}{(x)}
\nn \\ &&  \hs \left.
  + 61 \H{1, 1, 0, 0, 1}{(x)} + 3 \H{1, 1, 0, 1, 1}{(x)}
  + 3 \H{1, 1, 1, 0, 1}{(x)}\right)
\nn \\ &&  \hs
- \left(\frac{531}{2} L^2 + \frac{89}{4} \pi^2\right) \H{0, 0, 0, 1}{(x)}
- \left(\frac{311}{2} L^2 + \frac{619}{12}\pi^2\right) \H{0, 0, 1, 1}{(x)}
\nn \\ &&  \hs
- \left(\frac{247}{2} L^2 + \frac{307}{12} \pi^2\right) \H{0, 1, 0, 1}{(x)}
- \left(\frac{71}{2} L^2 + 32 \pi^2\right) \H{0, 1, 1, 1}{(x)}
\nn \\ &&  \hs
- \left(\frac{151}{2} L^2 - \frac{197}{12} \pi^2\right) \H{1, 0, 0, 1}{(x)}
- \left(\frac{107}{2} L^2 + 50 \pi^2\right) \H{1, 0, 1, 1}{(x)}
\nn \\ &&  \hs
- \left(\frac{35}{2} L^2 + 14 \pi^2\right) \H{1, 1, 0, 1}{(x)}
- \frac{3}{2}\left( L^2 +   \pi^2\right) \H{1, 1, 1, 1}{(x)}
\nn \\ &&  \hs
- \left(\frac{119}{2} L^3 + \frac{317}{12} \pi^2 L
- 455 \zeta_3 \right)\H{0, 0, 1}{(x)}
- \left(\frac{47}{2} L^3 + \frac{179}{12} \pi^2  L
- 120 \zeta_3 \right)\H{0, 1, 1}{(x)}
\nn \\ &&  \hs
- \left(\frac{35}{2} L^3 + \frac{35}{12} \pi^2  L
- 156 \zeta_3 \right)\H{1, 0, 1}{(x)}
- \left(\frac{3}{2} L^3 + \pi^2 L - 3 \zeta_3 \right)\H{1, 1, 1}{(x)}
\nn \\ &&  \hs
- \left(\frac{69}{8} L^4 + \frac{101}{8} \pi^2 L^2
- 291 \zeta_3 L + \frac{559}{90} \pi^4 \right)\H{0, 1}{(x)}
\nn \\ &&  \hs
- \left(\frac{9}{8} L^4 + \frac{25}{8} \pi^2 L^2 - 58 \zeta_3 L
+ \frac{13}{8} \pi^4\right) \H{1, 1}{(x)}
\nn \\ &&  \hs
- \left(\frac{27}{40} L^5 + \frac{25}{8} \pi^2 L^3
- \frac{183}{2} \zeta_3  L^2
   + \frac{131}{60} \pi^4 L - \frac{37}{12} \pi^2 \zeta_3
   + 57 \zeta_5 \right) \H{1}{(x)}
\nn \\ &&  \hs
+ \left(\frac{223}{12}  \pi^2 \zeta_3  + 149 \zeta_5 \right) L
+ \frac{167}{9} \zeta_3^2 - \frac{624607}{544320} \pi^6
\,. \label{TripleBoxValueA}
\end{eqnarray}

The result for the second triple box, defined in
\eqn{ThreeLoopIntegralB}
and shown in \fig{TripleBoxFigure}(b), is
\begin{equation}
I_4^{\threeloop b}(s,t) =
-\frac{1}{(-s)^{1+3\ep}\, t^2} \;
 \sum_{j=0}^6 \frac{ c_j(x,L)}{\ep^j}\;,
\label{ThreeLoopbResult}
\end{equation}
%
%
\def\hs{\hspace*{-84mm}}
where
\begin{eqnarray}
\;\;\; c_6=
{16\over 9}
\,, \;\;\;
%
%
c_5=
{13\over 6} L
\,, \;\;\;
%
%
c_4=
{1\over 2}L^2 - {19\over 12} \pi^2
\,,
\nn \\ &&  \hspace*{-84mm}
c_3=
{5\over 2} \left[\H{0, 0, 1}{(x)} + L \H{0, 1}{(x)}\right]
+ {5\over 4} \left[L^2 + \pi^2\right] \H{1}{(x)}
\nn \\ &&  \hs
- {7\over 12} L^3
- {157\over 72} L \pi^2 - {241\over 18} \zeta_3
\,,
\nn \\ &&  \hspace*{-84mm}
c_2=
 {1\over 2} \left[11 \H{0, 0, 0, 1}{(x)} - 5 \H{0, 0, 1, 1}{(x)}
 - 5 \H{0, 1, 0, 1}{(x)} - 5 \H{1, 0, 0, 1}{(x)}\right]
\nn \\ &&  \hs
+ {1\over 2}  L {} \left[14 \H{0, 0, 1}{(x)} -
5 \H{0, 1, 1}{(x)} - 5 \H{1, 0, 1}{(x)}\right]
+ {1\over 4}  L^2 \left[17 \H{0, 1}{(x)} -
5 \H{1, 1}{(x)}\right]
\nn \\ &&  \hs
+ {4\over 3} \pi^2 \H{0, 1}{(x)}
- {5\over 4} \pi^2 \H{1, 1}{(x)}
+ {5\over 3} L^3 \H{1}{(x)} + {25\over 12} L \pi^2 \H{1}{(x)}
\nn \\ &&  \hs
- {41\over 3} L \zeta_3
+ {5\over 2} \H{1}{(x)} \zeta_3 - {1\over 3}L^4
- {1\over 4}L^2 \pi^2 + {2429\over 6480} \pi^4
\,,
\nn \\ &&  \hspace*{-84mm}
c_1=
{1\over 2} \left[-55 \H{0, 0, 0, 0, 1}{(x)} - 59 \H{0, 0, 0, 1, 1}{(x)} -
31 \H{0, 0, 1, 0, 1}{(x)} + 5 \H{0, 0, 1, 1, 1}{(x)} \right.
\nn \\ &&  \hs
- 3 \H{0, 1, 0, 0, 1}{(x)} + 5 \H{0, 1, 0, 1, 1}{(x)} +
5 \H{0, 1, 1, 0, 1}{(x)} + 25 \H{1, 0, 0, 0, 1}{(x)}
\nn \\ &&  \hs
\left.
+ 5 \H{1, 0, 0, 1, 1}{(x)} + 5 \H{1, 0, 1, 0, 1}{(x)} +
5 \H{1, 1, 0, 0, 1}{(x)}\right]
\nn \\ &&  \hs
+ {1\over 2} L {}\left[22 \H{0, 0, 0, 1}{(x)} -
46 \H{0, 0, 1, 1}{(x)} - 18 \H{0, 1, 0, 1}{(x)} + 5 \H{0, 1, 1, 1}{(x)}
\right.
\nn \\ &&  \hs \left.
+ 10 \H{1, 0, 0, 1}{(x)} + 5 \H{1, 0, 1, 1}{(x)} + 5 \H{1, 1, 0, 1}{(x)}\right]
\nn \\ &&  \hs
+ {1\over 4} L^2 \left[64 \H{0, 0, 1}{(x)} - 33 \H{0, 1, 1}{(x)}
- 5 \H{1, 0, 1}{(x)} + 5 \H{1, 1, 1}{(x)}\right]
\nn \\ &&  \hs
+ {1\over 24} \pi^2 \left[25 \H{0, 0, 1}{(x)} -
128 \H{0, 1, 1}{(x)} + 40 \H{1, 0, 1}{(x)} + 30 \H{1, 1, 1}{(x)}\right]
\nn \\ &&  \hs
+ {1\over 12} L^3 \left[71 \H{0, 1}{(x)} - 20 \H{1, 1}{(x)}\right]
\nn \\ &&  \hs
+ {1\over 24} L \pi^2\left[153 \H{0, 1}{(x)} - 50 \H{1, 1}{(x)}\right]
+ {1\over 2} \left[8 \H{0, 1}{(x)}
- 5 \H{1, 1}{(x)}\right] \zeta_3
\nn \\ &&  \hs
+ {43\over 48}L^4 \H{1}{(x)} + {71\over 48} L^2 \pi^2 \H{1}{(x)}
- {5\over 144} \pi^4 \H{1}{(x)} - {5\over 2} L \H{1}{(x)} \zeta_3
+ {7\over 48} L^5
\nn \\ &&  \hs
+ {227\over 144} L^3 \pi^2 + {13\over 4} L^2 \zeta_3
+ {10913\over 8640} L \pi^4
+ {3257\over 216} \pi^2 \zeta_3 - {889\over 10} \zeta_5
\,,
\nn \\ &&  \hspace*{-84mm}
c_0=
{1\over 2} \left[379 \H{0, 0, 0, 0, 0, 1}{(x)} + 343 \H{0, 0, 0, 0, 1, 1}{(x)} +
419 \H{0, 0, 0, 1, 0, 1}{(x)} + 347 \H{0, 0, 0, 1, 1, 1}{(x)}
\right.
\nn \\ &&  \hs
+ 355 \H{0, 0, 1, 0, 0, 1}{(x)} + 175 \H{0, 0, 1, 0, 1, 1}{(x)} +
223 \H{0, 0, 1, 1, 0, 1}{(x)} - 5 \H{0, 0, 1, 1, 1, 1}{(x)}
\nn \\ &&  \hs
+ 151 \H{0, 1, 0, 0, 0, 1}{(x)} + 3 \H{0, 1, 0, 0, 1, 1}{(x)} +
51 \H{0, 1, 0, 1, 0, 1}{(x)} - 5 \H{0, 1, 0, 1, 1, 1}{(x)}
\nn \\ &&  \hs
+ 99 \H{0, 1, 1, 0, 0, 1}{(x)} - 5 \H{0, 1, 1, 0, 1, 1}{(x)} -
5 \H{0, 1, 1, 1, 0, 1}{(x)} - 193 \H{1, 0, 0, 0, 0, 1}{(x)}
\nn \\ &&  \hs
- 169 \H{1, 0, 0, 0, 1, 1}{(x)} - 121 \H{1, 0, 0, 1, 0, 1}{(x)} -
5 \H{1, 0, 0, 1, 1, 1}{(x)} - 73 \H{1, 0, 1, 0, 0, 1}{(x)}
- 5 \H{1, 0, 1, 0, 1, 1}{(x)}
\nn \\ &&  \hs
\left.
 - 5 \H{1, 0, 1, 1, 0, 1}{(x)} -
25 \H{1, 1, 0, 0, 0, 1}{(x)} - 5 \H{1, 1, 0, 0, 1, 1}{(x)} -
5 \H{1, 1, 0, 1, 0, 1}{(x)} - 5 \H{1, 1, 1, 0, 0, 1}{(x)}\right]
\nn \\ &&  \hs
+ {1\over 2} L {}\left[98 \H{0, 0, 0, 0, 1}{(x)} - 22 \H{0, 0, 0, 1, 1}{(x)} +
98 \H{0, 0, 1, 0, 1}{(x)} + 238 \H{0, 0, 1, 1, 1}{(x)}
+ 78 \H{0, 1, 0, 0, 1}{(x)}
\right.
\nn \\ &&  \hs
 + 66 \H{0, 1, 0, 1, 1}{(x)} +
114 \H{0, 1, 1, 0, 1}{(x)} - 5 \H{0, 1, 1, 1, 1}{(x)} -
82 \H{1, 0, 0, 0, 1}{(x)} - 106 \H{1, 0, 0, 1, 1}{(x)}
\nn \\ &&  \hs \left.
- 58 \H{1, 0, 1, 0, 1}{(x)} - 5 \H{1, 0, 1, 1, 1}{(x)} -
10 \H{1, 1, 0, 0, 1}{(x)} - 5 \H{1, 1, 0, 1, 1}{(x)} -
5 \H{1, 1, 1, 0, 1}{(x)}\right]
\nn \\ &&  \hs
+ {1\over 4}L^2 \left[124 \H{0, 0, 0, 1}{(x)} -
208 \H{0, 0, 1, 1}{(x)} - 44 \H{0, 1, 0, 1}{(x)} + 129 \H{0, 1, 1, 1}{(x)}
\right.
\nn \\ &&  \hs \left.
- 20 \H{1, 0, 0, 1}{(x)} - 43 \H{1, 0, 1, 1}{(x)} + 5 \H{1, 1, 0, 1}{(x)} -
5 \H{1, 1, 1, 1}{(x)}\right]
\nn \\ &&  \hs
+ {1\over 24} \pi^2 \left[183 \H{0, 0, 0, 1}{(x)} -
121 \H{0, 0, 1, 1}{(x)} + 375 \H{0, 1, 0, 1}{(x)} + 704 \H{0, 1, 1, 1}{(x)}
\right.
\nn \\ &&  \hs \left.
+ 31 \H{1, 0, 0, 1}{(x)} - 328 \H{1, 0, 1, 1}{(x)} - 40 \H{1, 1, 0, 1}{(x)} -
30 \H{1, 1, 1, 1}{(x)}\right]
\nn \\ &&  \hs
+ {1\over 12} L^3 \left[260 \H{0, 0, 1}{(x)}
- 215 \H{0, 1, 1}{(x)} - 7 \H{1, 0, 1}{(x)} + 20 \H{1, 1, 1}{(x)}\right]
\nn \\ &&  \hs
+ {1\over 24} L \pi^2 \left[326 \H{0, 0, 1}{(x)}- 633 \H{0, 1, 1}{(x)}
+ 127 \H{1, 0, 1}{(x)} + 50 \H{1, 1, 1}{(x)}\right]
\nn \\ &&  \hs
- {1\over 2}  \left[- 3 L \H{0, 1}{(x)}
- 5 L \H{1, 1}{(x)} + 165 \H{0, 0, 1}{(x)}
+ 104 \H{0, 1, 1}{(x)} -
68 \H{1, 0, 1}{(x)} - 5 \H{1, 1, 1}{(x)}\right] \zeta_3
\nn \\ &&  \hs
+ {1\over 48} L^4\left[309 \H{0, 1}{(x)} - 43 \H{1, 1}{(x)}\right]
+ {1\over 48} L^2 \pi^2\left[725 \H{0, 1}{(x)} - 71 \H{1, 1}{(x)}\right]
\nn \\ &&  \hs
+ {1\over 720} \pi^4 \left[1848 \H{0, 1}{(x)} + 25 \H{1, 1}{(x)}\right]
\nn \\ &&  \hs
+ {37\over 120} L^5 \H{1}{(x)}
+ {11\over 8} L^3 \pi^2 \H{1}{(x)} + {641\over 720} L \pi^4 \H{1}{(x)}
+ {38\over 3} L^3 \zeta_3 + {479\over 18} L \pi^2 \zeta_3
\nn \\ &&  \hs
- 2 L^2 \H{1}{(x)} \zeta_3
- {269\over 24} \pi^2 \H{1}{(x)} \zeta_3 + {129\over 2} \H{1}{(x)} \zeta_5
+ {151\over 720}L^6 + {373\over 288} L^4 \pi^2
\nn \\ &&  \hs
+ {3163\over 2880} L^2 \pi^4 - {1054\over 5} L \zeta_5
+ {1391417\over 3265920} \pi^6 + {197\over 6} \zeta_3^2
\,. \label{TripleBoxValueB}
\end{eqnarray}


\end{document}